%% file: paper.tex
\documentclass[10pt, conference]{IEEEtran}
\usepackage[sort&compress, numbers]{natbib}
\usepackage{verbatim}
\usepackage{caption}
\usepackage{subcaption}
\usepackage[]{algorithm2e}
\usepackage{graphicx}
\usepackage{booktabs}
\usepackage{relsize}
\usepackage{colortbl}
\usepackage{enumerate}
\usepackage{amsmath}
\usepackage{tikz}
\usepackage{pgf}
\usepackage{pbox}
\usepackage{rotating}
\usepackage{multirow}
\usepackage{balance}
\usepackage{tablefootnote}
\usepackage[bookmarks=false]{hyperref}
\usepackage{flushend}

\usepackage{cleveref}[2012/02/15]
\usepackage{arydshln}
\crefformat{footnote}{#2\footnotemark[#1]#3}

\usepackage{soul}

\newcommand{\startlist}{\begin{list}{\labelitemi}{\leftmargin=1em}\setlength{\itemsep}{-1mm}}
\newcommand{\stoplist}{\end{list}}

\newcommand{\smallsection}[1]{\noindent \underline{#1}.}
\newcommand{\ea}{{\em et al.}}

\usepackage{tcolorbox}
\newcommand{\rqi}{Does our \name~\underline{outperform} the state-of-the-art JIT defect prediction approaches?}
\newcommand{\rqii}{Is our \name~more \underline{cost-effective} than the  state-of-the-art JIT defect prediction approaches?}
\newcommand{\rqiii}{Is our \name~\underline{faster} than the state-of-the-art JIT defect prediction approaches?}
\newcommand{\rqiiii}{How effective is our \name~for prioritizing  defective \underline{lines} of a given defect-introducing commit?}

\newcommand{\name}{JITLine}

\begin{document}

% Prof. Kla suggested title "Towards Line-Level Just-In-Time Defect Prediction" but I want readers know that here we use LIME to make explanation of model so I added "LIME-JIT"
\title{\name: A Simpler, Better, Faster, Finer-grained Just-In-Time Defect Prediction}

\author{%
    \IEEEauthorblockN{Chanathip Pornprasit, Chakkrit (Kla) Tantithamthavorn}
    \IEEEauthorblockA{Monash University, Melbourne, Australia.}
}

\maketitle

\begin{abstract} 
A Just-In-Time (JIT) defect prediction model is a classifier to predict if a commit is defect-introducing.
Recently, CC2Vec---a deep learning approach for Just-In-Time defect prediction---has been proposed.
However, CC2Vec requires the whole dataset (i.e., training + testing) for model training, assuming that all unlabelled testing datasets would be available beforehand, which does not follow the key principles of just-in-time defect predictions.
Our replication study shows that, after excluding the testing dataset for model training, the F-measure of CC2Vec is decreased by 38.5\% for OpenStack and 45.7\% for Qt, highlighting the negative impact of excluding the testing dataset for Just-In-Time defect prediction.
In addition,
CC2Vec cannot perform fine-grained predictions at the line level (i.e., which lines are most risky for a given commit).

In this paper, we propose \name---a Just-In-Time defect prediction approach for predicting defect-introducing commits and identifying lines that are associated with that defect-introducing commit (i.e., defective lines).
Through a case study of 37,524 commits from OpenStack and Qt, we find that our \name~approach is at least 26\%-38\% more accurate (F-measure), 17\%-51\% more cost-effective (PCI@20\%LOC), 70-100 times faster than the state-of-the-art approaches (i.e., CC2Vec and DeepJIT) and the fine-grained predictions at the line level by our approach are 133\%-150\% more accurate (Top-10 Accuracy) than the baseline NLP approach.
Therefore, our \name~approach may help practitioners to better prioritize defect-introducing commits and better identify defective lines.
\end{abstract}

\input{sections/introduction.tex}

\input{sections/background.tex}

\input{sections/approach.tex}

\input{sections/experiment.tex}

\input{sections/results.tex}
\input{sections/discussion}
% \input{sections/threats.tex}
% \input{sections/relatedwork.tex}

\section{Conclusions}\label{sec:conclusions}

In this paper, we propose \name~approach, a machine learning-based JIT defect approach for predicting defect-introducing commits and identifying defective lines that are associated with that commit.
Then, we conduct our empirical study to demonstrate that our \name~approach is better (RQ1), more cost-effective (RQ2), faster (RQ3) and more fine-grained (RQ4) than the state-of-the-art JIT defect prediction approaches (i.e., EARL, DeepJIT, and CC2Vec).

Therefore, our \name~approach may help practitioners to better prioritize defect-introducing commits and better identify defective lines.
In addition, our results highlight the negative impact of excluding testing datasets in model training and the importance of exploring simple solutions (e.g., explainable AI approaches) first over complex and compute-intensive deep learning approaches.

% We first evaluate the performance of bug-introducing commit prediction on deep learning-based approaches (i.e., DeepJIT and CC2Vec) and other traditional approaches (i.e., EALR and our \name).
% In addition, we evaluate these approaches with respect to effort-aware measure.
% From the experiment result, our approach outperforms the baselines in term of AUC, precision, F1 and FAR, while achieving the greatest cost-effectiveness when considering PCI@20\%LOC, Effort@20\%Recall and P\textsuperscript{Opt}.
% Furthermore, we show that the training time of \name~is faster than the deep learning baselines, which are trained on both CPU and GPU.
% At the end, we illustrate that our approach outperforms N-gram model for priortizing defective lines of defect-introducing commits.

% \section*{Acknowledgments}

\textbf{Acknowledgement.} Chakkrit Tantithamthavorn was supported by ARC DECRA Fellowship (DE200100941).

\bibliographystyle{IEEEtranS}
\bibliography{myref,additionalref}

% \newpage
% .
% \newpage

% \section{Appendix}
% \input{sections/appendix}

% \usepackage{multirow}
% \usepackage{graphicx}

\end{document}

%% file: sections/introduction.tex
%%%%%%%%%%%%%%%%%%%%%%%%%%%%%%%%
\section{Introduction}
%%%%%%%%%%%%%%%%%%%%%%%%%%%%%%%%

Modern software development cycles tend to release software products in a short-term period.
% ~\cite{adams2016modern}.
Such short-term software development cycles often pose critical challenges to modern Software Quality Assurance (SQA) practices. 
Therefore, continuous code quality tools (e.g., CI/CD, modern code review, static analysis) have been heavily adopted to early detect software defects.
However, SQA teams cannot effectively inspect every commit given limited SQA resources.

Just-in-time (JIT) defect prediction~\cite{Kamei2010, kim2007predicting} is proposed to predict if a commit will introduce defects in the future.
Such commit-level predictions are useful to help practitioners prioritize their limited SQA resources on the most risky commits during the software development process.
In the past decades, several machine learning approaches are employed for developing JIT defect prediction models~\cite{Kim2008, Shivaji2013, Fukushima2014, Rajbahadur2017}.
% , e.g., Random Forests (RF), or Logistic Regression (LR).
However, these approaches often rely on handcrafted commit-level features (e.g., Churn).

Recently, several deep learning approaches have been proposed for Just-In-Time defect prediction (e.g., DeepJIT~\cite{hoang2019deepjit} and CC2Vec~\cite{CC2Vec}).
Hoang~\ea~\cite{CC2Vec} found that their CC2Vec approach outperforms DeepJIT for Just-In-Time defect prediction.
CC2Vec requires both training and unlabelled testing datasets for training CC2Vec models, assuming that all unlabelled testing datasets would be available beforehand.
However, these assumptions of CC2Vec do not follow the key principles of the Just-In-Time defect prediction: 
(1) the predictions of the CC2Vec approach cannot be made immediately for a newly arrived commit; and (2) it is unlikely that the unlabelled testing dataset would be available beforehand when training JIT models.
Thus, we perform a replication study to confirm the merit of previous experimental findings and extend their experiment by excluding testing datasets and evaluate with five additional evaluation measures.

% Thus, we first conduct a Replication Study (RS) of the state-of-the-art deep learning approach (CC2Vec~\cite{CC2Vec}) for JIT defect prediction and extend their experiment with a realistic model training scenario (i.e., excluding testing datasets) with five additional measures.

% However, CC2Vec requires the whole dataset (i.e., training+testing) for model training, which does not mimic the realistic model training scenario.
% However, Hoang~\ea~\cite{CC2Vec}'s approach relies on the whole dataset when training their CC2Vec models.
% Yet, .
% While the replicability and reproducibility of deep learning approaches are known to be challenging for SE tasks~\cite{liu2020replicability}, there exists no replication study of Hoang~\ea~\cite{CC2Vec} for Just-In-Time defect prediction.

\begin{enumerate}[{\bf RS1)}]
\item {\bf Can we replicate the results of deep learning approaches for Just-In-Time defect prediction?}\\
Similar to the original study~\cite{CC2Vec}, we are able to replicate the results of CC2Vec.
\item {\bf How does CC2Vec perform for Just-In-Time defect prediction after excluding testing datasets?}\\
After excluding testing datasets when developing the JIT models, we find that the F-measure of CC2Vec is decreased by 38.5\% for OpenStack and 45.7\% for Qt.
In addition, CC2Vec achieves a high False Alarm Rate (FAR) of 0.87 for OpenStack and 0.63 for Qt, indicating that 63\%-87\% clean commits are incorrectly predicted as defect-introducing.
Thus, developers still waste many unnecessarily effort to inspect clean commits that are incorrectly predicted as defect-introducing.
\end{enumerate}

In addition, Hoang~\ea~\cite{CC2Vec} did not compare their approach with simple JIT approaches, did not evaluate the cost-effectiveness, did not report the computational time, and cannot perform fine-grained predictions at the line level.
Thus, it remains unclear about the practical value of the CC2Vec approach when considering the amount of effort that developers need to inspect.

In this paper, we propose \name---a machine learning-based Just-In-Time defect prediction approach that can both predict defect-introducing commits and identify defective lines that are associated with that commit.
We evaluate our \name~approach with the state-of-the-art commit-level JIT defect prediction approaches (i.e., EARL~\cite{Kamei2013}, DeepJIT~\cite{hoang2019deepjit}, and CC2Vec~\cite{CC2Vec}) with respect to six traditional measures (i.e, AUC, F-measure, False Alarm Rate, Distance-to-Heaven, Precision, and Recall), three cost-effectiveness measures (i.e., PCI@20\%LOC, Effort@20\%Recall, P\textsubscript{Opt}).
In addition, we also compare our approach with a baseline line-level JIT defect localization by Yan~\cite{yan2020just} using four line-level effort-aware measures (i.e., Top-10 Accuracy, Recall@20\%LOC, Effort@20\%Recall\textsubscript{line}, Initial False Alarm). 
Through a case study of 37,524 total commits that span across two large-scale open-source software projects (i.e., OpenStack and Qt), 
we address the following four research questions:

\begin{enumerate}[{\bf RQ1)}]

\item {\bf \rqi}\\
% Our \name~achieves an F-measure 390\%-475\% higher and an AUC 28\%-73\% higher than the state-of-the-art approaches.
Our \name~approach achieves F-measure 26\%-38\% higher than the state-of-the-art approaches (i.e., CC2Vec).
Our \name~achieves a False Alarm Rate (FAR) 94\%-97\% lower than the CC2Vec approach.

\item {\bf \rqii}\\
Our \name~is 17\%-51\% more cost-effective than the state-of-the-art approaches in term of PCI\@20\%Effort.
In addition, our \name~can save the amount of effort by 89\%-96\% to find the same number of actual defect-introducing commits (i.e., 20\% Recall) when compared to the state-of-the-art approaches.

\item {\bf \rqiii}\\
Our \name~is 70-100 times faster than the deep learning approaches for Just-In-Time defect prediction.

\item {\bf \rqiiii}\\
Our \name~approach is 133\%-150\% more accurate than the baseline approach by Yan~\ea~\cite{yan2020just} for identifying actual defective lines in the top-10 recommendations.
Our \name~approach requires 17\%-27\% less amount of effort than the baseline approach in order to find the same amount of actual defective lines.

\end{enumerate}

\smallsection{Contributions} The contributions of this paper are as follows:

% \begin{itemize}
%     \item \todo{please list the contributions here}
%     \item A replication study of XXX.
%     \item propose a new technique. improve, 
    
%     \item We collect line-level ground-truth of defective code change in Openstack and Qt.
%     \item We adapt existing machine learning technique and explainable AI tool (LIME) to build line-level JIT defect prediction
%     \item We evaluate the ability to predict defective commit of our model against state-of-the-art deep learning approaches.
%     \item We compare our line-level ranking performance with ngram model.
%     \item We provide a replication package to foster the future reprehensibility (in the supplementary materials due to a double-blind policy). Final files will be available in a public archive \url{http://??}. 
% \end{itemize}

\begin{itemize}
    \item We conduct a replication study of the state-of-the-art deep learning approach (CC2Vec~\cite{CC2Vec}) for JIT defect prediction and extend their experiment by excluding testing datasets with five evaluation measures (Section~\ref{sec:revisiting}). 
    \item We propose \name---a machine learning-based Just-In-Time defect prediction approach that can both predict defect-introducing commits and identify their associated defective lines (Section~\ref{sec:approach}).
    \item We evaluate our \name~approach at the commit level with the state-of-the-art JIT defect prediction approaches with respect to six traditional measures, three cost-effectiveness measures, and at the line level with four effort-aware line-level measures (Section~\ref{sec:results}).
    \item Our results show that our \name~approach outperforms (RQ1), more cost-effective (RQ2), faster (RQ3), and more fine-grained (RQ4) than the state-of-the-art approaches.
    % along with four dimensions, i.e., accuracy, cost-effectiveness, computational time, and the effectiveness at line level.
    % we create line-level JIT defect prediction on Openstack and Qt dataset
    % \item we perform experiment on commit-level defect prediction by using random forest classifier
    % \item we evaluate performance of model on commit-level defect classification task based on AUC score
    % \item we perform experiment on line-level defect prediction by using effort-aware measure as metric
    % \item + ground-truth datasets for line-level just-in-time defect prediction
\end{itemize}

% \smallsection{Paper organization}
% Section~\ref{sec:background} situates this paper with respect to the related work.
% Section~\ref{sec:design} discusses the design of our case study, while Section~\ref{sec:results} presents the results with respect to our two research questions.
% Section~\ref{sec:threats} discloses the threats to the validity of our study.
% Finally, Section~\ref{sec:conclusions} draws conclusions.

%% file: sections/background.tex
\begin{table*}[t]
\centering
\caption{The results of our replication study of CC2Vec~\cite{CC2Vec} when using ``train+test'' and ``train only'' for model training.}
\label{tab:replication-study}
\begin{tabular}{lc|c|c|c|c|c|c|c|c|c|c|c|c|}
\cline{3-14}
    & \multicolumn{1}{l|}{} & \multicolumn{6}{c|}{\textbf{OpenStack}}                                                         & \multicolumn{6}{c|}{\textbf{Qt}}                                                                
    \\ \cline{3-14} 
    & \multicolumn{1}{l|}{} & \textbf{AUC} & \textbf{F1} & \textbf{FAR} & \textbf{d2h} & \textbf{Precision} & \textbf{Recall} & \textbf{AUC} & \textbf{F1} & \textbf{FAR} & \textbf{d2h} & \textbf{Precision} & \textbf{Recall} 
    \\ \hline
\multicolumn{1}{|c|}{\multirow{2}{*}{\textbf{CC2Vec {[}Train+Test{]}}}} & \textbf{Original}     & \cellcolor[HTML]{32CB00}0.81         & -           & -            & -            & -                  & -               & \cellcolor[HTML]{32CB00}0.82         & -           & -            & -            & -                  & -               
\\ \cline{2-14} 
\multicolumn{1}{|c|}{}                                         & \textbf{Ours}         & \cellcolor[HTML]{32CB00}0.80         & \cellcolor[HTML]{FD6864}0.39        & 0.26         & 0.28         & 0.27               & 0.70            & \cellcolor[HTML]{32CB00}0.84         & \cellcolor[HTML]{FD6864}0.35        & 0.17         & 0.25         & 0.24               & 0.70            
\\ \hline
\multicolumn{1}{|l|}{\textbf{CC2Vec {[}Train Only{]}}}                  & \textbf{Ours}         & 0.77         & \cellcolor[HTML]{FD6864}0.24        & 0.87         & 0.61         & 0.14               & 0.99            & 0.81         & \cellcolor[HTML]{FD6864}0.19        & 0.63         & 0.45         & 0.10               & 0.96           
\\ \hline
\end{tabular}
\end{table*}

%%%%%%%%%%%%%%%%%%%%%%%%%%%%%%%%
\section{Background}\label{sec:background}

% In this section, we provide the background of the definitions of commits and the state-of-the-art approaches for Just-In-Time defect prediction.

Commits created by developers are often used to describe new features, bug fixes, refactoring, etc. 
One commit contains three main pieces of information, i.e., a commit message, a code change, and their meta-data information (e.g., churn, author name). 
The commit message is used to describe the semantics of the code changes, while the code change indicates changed lines (i.e., added/modified/deleted lines).

In large-scale software projects, there is a stream of commits that developers need to review and inspect. 
However, due to the limited SQA resources, Just-In-Time defect prediction approaches have been proposed to help developers prioritize their limited SQA resources on the most risky commits~\cite{Kamei2013,Kim2008}.
Below, we discuss three state-of-the-art approaches for Just-In-Time defect prediction.

\emph{EALR}~\cite{Kamei2013} is an Effort-Aware JIT defect prediction method using a Logistic Regression model with traditional commit-level software metrics (e.g., churn).
% EALR trains a logistic regression model .
EALR generates a rank of defect-introducing commits by considering the amount of inspection effort---i.e., the predicted probability is normalized by the commit size (i.e., churn).
However, such techniques often rely on handcrafted feature engineering.

\emph{DeepJIT}~\cite{hoang2019deepjit} is an end-to-end deep learning framework for Just-in-Time defect prediction.
DeepJIT automatically generates features using a Convolutional Neural Network (CNN) architecture.
Generally, DeepJIT takes the commit message and the code change as an input into two CNN models in order to generate a vector representation---i.e, one CNN for generating commit message vectors and another CNN for generating code changes vectors.
Finally, the concatenation of both the commit message vector and the code change vector is input into the fully-connected layer to generate the probability of defect-introducing commit.

% In depth, this approach begins by transforming commit message and code change to .
% Then, two Convolutional Neural Networks (CNNs) \cite{CNN} are employed, one for extracting features from commit message, another one for extracting features from code change.
% The result of both CNNs are two feature vectors, which are concatenated to form the final feature vector.
% This vector is then used as input for fully-connected layer to produce probability of bug-introducing commit.

% \begin{figure}[t]
% \centering
% \includegraphics[width=\columnwidth]{figures/jit.pdf}
% \caption{The Just-In-Time Defect Prediction.}
% \label{fig:jit}
% \end{figure}

\emph{CC2Vec}~\cite{CC2Vec} is an approach to learn the distributed representation of commit.
Traditionally, one commit has a hierarchical structure--i.e., one commit consists of changed files, one change file consists of changed hunks, one change hunk consists of changed lines, one changed line consists of changed tokens.
Unlike DeepJIT that ignores the information about the hierarchical structure of code commits, CC2Vec has been proposed to automatically learn the hierarchical structure of code commits using a Hierarchical Attention Network (HAN) architecture.
The goal of CC2Vec is to learn the relationship between the actual code changes and the semantic of that code changes (i.e., the first line of commit messages).
Then, in the feature extraction layer, HAN is used to build vector representations of changed lines; these vectors are then used to construct vector representations of hunks; and then these vectors are aggregated to construct the embedding vector of the removed or added code. 
Then, the embedding vectors of the removed code and added code is input into a fully-connected layer to generate a vector that represents the code change.

Recently, Hoang~\ea~has shown that the combination of CC2Vec and DeepJIT outperforms the stand-alone DeepJIT approach.
In particular, they used CC2Vec to generate a vector representation of code changes.
Then, such code changes vector is concatenated with the commit message vectors and the code change vectors that are generated by DeepJIT to generate a final vector representation.
Finally, the concatenation vector is input into the fully-connected layer to generate the probability of defect-introducing commit.

\section{A Replication Study of the State-of-the-art Deep Learning Approach for JIT Defect Prediction}\label{sec:revisiting}

In this section, we present the motivation, approach, and results of our replication study (RS) of CC2Vec for Just-In-Time defect prediction.

\begin{figure}[t]
\centering
\includegraphics[width=\columnwidth]{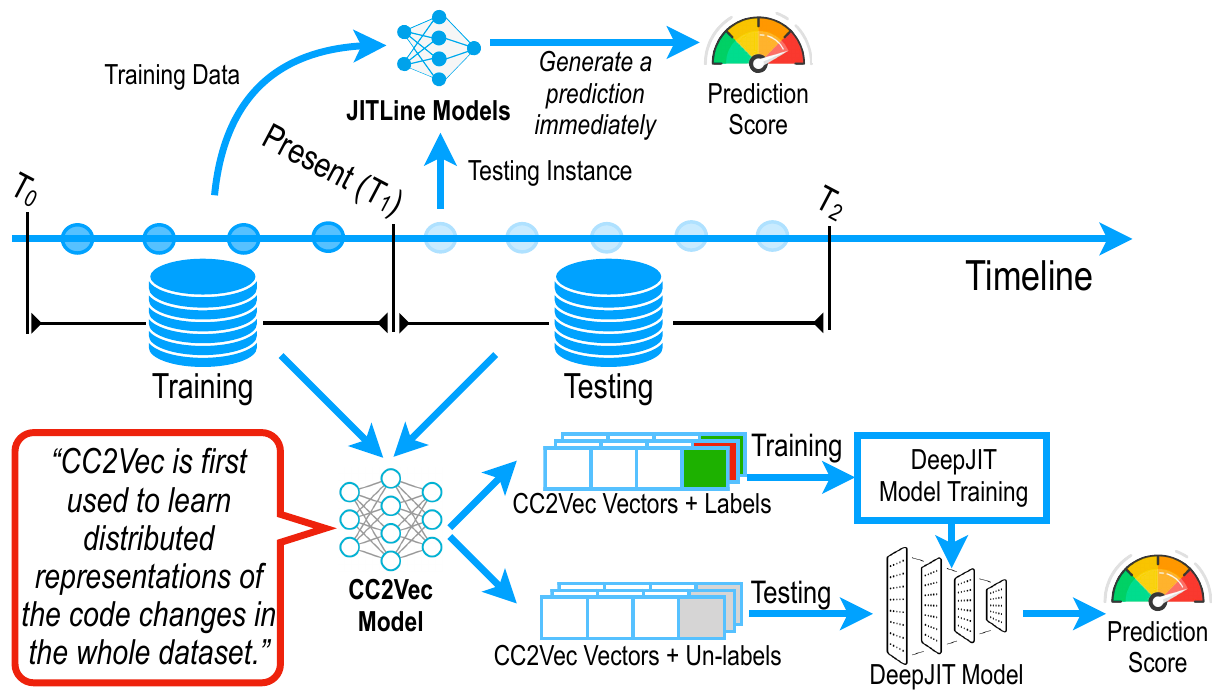}
\caption{The comparison between the workflow of JITLine that can immediately generate predictions and the workflow of CC2Vec+DeepJIT~\cite{CC2Vec} which requires testing dataset to be available beforehand for training CC2Vec+DeepJIT models.}
\label{fig:cc2vec}
\end{figure}

\textbf{Motivation.}
One of the key principles of \emph{Just-In-Time} defect prediction models is to \emph{generate predictions as soon as possible} for a newly arrived commit.
Let's consider $T_1$ as the present (see Figure \ref{fig:cc2vec}), the whole historical data will be used for training a JIT model in order to immediately generate a prediction of a newly arrived commit.
However, CC2Vec requires both training and unlabelled testing datasets for training CC2Vec models (i.e., the periods of $T_0$-$T_1$ and $T_1$-$T_2$), assuming that all unlabelled testing datasets would be available beforehand.
In particular, Hoang~\ea~ (Section 3.3.3 of the original study~\cite{CC2Vec}) stated that \emph{``CC2Vec is first used to learn distributed representations of the code changes in the whole dataset.
All patches from the training and testing dataset are used since the log messages of the testing dataset are not part of the predictions of the task''}.
This indicates that the unlabelled testing dataset needs to be available beforehand for training CC2Vec models.
However, these assumptions of CC2Vec do not follow the key principles of the Just-In-Time defect prediction: 
(1) the predictions of the CC2Vec approach cannot be  made immediately for a newly arrived commit; and (2) it is unlikely that the unlabelled testing dataset would be available beforehand when training JIT models.
Thus, it remains unclear what the performance of CC2Vec for Just-In-Time defect prediction is after considering the key principle of Just-In-Time defect prediction (i.e., excluding testing dataset for model training).
In addition, several other performance measures (e.g., F-measure) have not been evaluated in the original study.
Thus, we (RS1) perform a replication study to confirm the merit of previous experimental findings and (RS2) extend their experiment by excluding testing datasets and evaluate with five additional evaluation measures.

\subsection*{\textbf{(RS1) Can we replicate the results of deep learning approaches for Just-In-Time defect prediction?}}

\smallsection{Approach}
To address RS1, we first download the replication package of Hoang~\ea~\cite{CC2Vec}.
% \footnote{\todo{zenodo?...}}
We carefully study the replication package to understand all details. 
Then, we execute the source code followed by the instructions and datasets provided by Hoang~\ea~\cite{CC2Vec}.
Finally, we compute a relative percentage between our results and the original paper as follows: $\% = (\frac{\mathrm{ours}-\mathrm{original}}{\mathrm{original}})\times 100\%$

\smallsection{Results}
\textbf{Similar to the original study~\cite{CC2Vec}, we are able to replicate the results of CC2Vec.}
Table \ref{tab:replication-study} (see the green cells) shows that, in our experiment, CC2Vec achieves an AUC of 0.80 for OpenStack and 0.84 for Qt, while the original paper reported an AUC of 0.81 for OpenStack and 0.82 for Qt.
Our results are only 1\%-2\% different when compared to the original paper.
This finding confirms that the results of CC2Vec are replicable for Just-In-Time defect prediction.

\subsection*{\textbf{(RS2) How does CC2Vec perform for Just-In-Time defect prediction after excluding testing datasets?}}

% \smallsection{Motivation} 

% % added by Oat
% In other words, both training and testing data are used to train CC2Vec because CC2Vec is not used directly to make a prediction.
% It is used to extract features from code change for another model training step.
% In addition, other evaluation measures have not been evaluated in the original paper~\cite{Hoang2019}.

\smallsection{Approach} To address RS2, we repeat the experiment of Hoang~\ea~\cite{CC2Vec} in two settings---i.e., the original experiment with training and testing datasets and our experiment with training datasets only.
In addition, we extend their experiment by evaluating the CC2Vec approach using five additional evaluation measures (i.e., F-measure, False Alarm Rate, Distance-to-Heaven, Precision, and Recall).
% The details of the evaluation measures are described in the Approach of RQ1 in Section~\ref{sec:results}.

\smallsection{Results} \textbf{After excluding testing datasets when developing the JIT models, we find that the F-measure of CC2Vec is decreased by 38.5\%(0.35$\rightarrow$0.19) for OpenStack. and 45.7\%(0.39$\rightarrow$0.24) for Qt.}
Table \ref{tab:replication-study} (see the red cells) shows the result between two experimental settings (i.e., training+testing vs. training only) with respect to AUC, F1, FAR, d2h, Precision and Recall.
We find that the values of several performance measures (i.e, AUC, F-measure, FAR, d2h) are negatively impacted by the exclusion of the testing datasets.
We find that AUC is decreased by 3.9\% for OpenStack and 3.7\% for Qt, while False Alarm Rates (FAR) are increased by 234.62\% for OpenStack and 270.59\% for Qt.
Similarly, the d2h value is increased by 126.92\% for Openstack and 80\% for Qt.
The higher FAR and d2h of CC2Vec has to do with the substantial increasing Recall to 0.99 for OpenStack and 0.96 for Qt---i.e., CC2Vec predicts most of the commits as defect-introducing (higher Recall), but many of the predictions are incorrect (higher FAR, less Precision).
These findings indicate that the exclusion of testing datasets in model training has a large negative impact on the performance of CC2Vec (i.e., producing higher False Alarm Rates).
Thus, developers have to waste unnecessarily effort on inspecting clean commits that are incorrectly predicted as defect-introducing.

\section{Related Work and Research Questions}

In this section, we discuss the following four main limitations of prior studies with respect to the literature in order to motivate our approach and research questions.

\textbf{First, several traditional machine learning-based JIT approaches have not been compared with the deep learning approaches for JIT defect prediction.}
Recently, researchers found that several simple approaches often outperform deep learning approaches in SE tasks.
For example, Hellendoorn~\cite{hellendoorn2017deep}, Fu and Menzies~\cite{fu2017easy}, Liu~\ea~\cite{liu2018neural}.
Menzies~\ea~\cite{menzies2018500+} suggested that researchers should explore simple and fast approaches before applying deep learning approaches on SE tasks.
However, Hoang~\ea~\cite{CC2Vec} did not compared their CC2Vec approach with other simple approaches (e.g., logistic regression and random forest).
Therefore, we wish to investigate if our approach outperforms the deep learning approaches for Just-In-Time defect prediction.

\textbf{Second, the cost-effectiveness of deep learning approaches for JIT defect prediction has not been investigated.}
Prior work pointed out that different code changes often require different amount of code inspection effort~\cite{mende2010effort,huang2017supervised}---i.e., large code changes often require a high amount of code inspection effort.
However, Hoang~\ea~\cite{CC2Vec} did not investigate the cost-effectiveness of their CC2Vec approach.
In addition, the CC2Vec approach does not take into consideration the effort required to inspect code changes when prioritizing defect-introducing commits.
Therefore, we wish to investigate if our approach is more cost-effective than the deep learning approaches for Just-In-Time defect prediction.

\textbf{Third, the computational time of deep learning approaches JIT defect prediction has not been investigated.}
Several researchers raised concerns that deep learning approaches are often complex and very expensive in terms of GPU costs/CPU time.
For example,  Jiang~\ea~\cite{jiang2017automatically}'s approach requires 38 hours for training their deep learning models on NVIDIA GeForce GTX 1070.
Menzies~\ea~\cite{menzies2018500+} found that a simple approach that is 500+ times faster achieves similar performance to deep learning approaches. 
Therefore, we wish to investigate if our approach is faster than the deep learning approaches for Just-In-Time defect prediction.

% \textbf{Finally, there exists no machine learning approaches for Just-In-Time defect prediction that can perform fine-grained predictions at the line level.}
\textbf{Finally, there exists no machine learning approaches for fine-grained Just-In-Time defect prediction at line level.}
Recently, Pascarella~\ea~\cite{PASCARELLA2019} proposed a fine-grained JIT defect prediction model which based on handcrafted features to prioritize which changed files in a commit should be review first.
However, this approach  cannot identify defective lines of the changed files.
Recently, Yan~\ea~\cite{yan2020just} proposed a fine-grained JIT defect localization at the line level to help developers to locate and address defects using less effort.
Yan~\ea~\cite{yan2020just} proposed a two-phase approach---i.e., the ML model trained on software metrics (e.g., \#added\_lines) is first used to identify which commits are the most risky, then the N-gram model trained on textual features is finally used to localise the riskiest lines.
On the other hand, a recent work by Wattanakriengkrai~\ea~\cite{wattanakriengkrai2020predicting} pointed out that a machine learning approach outperforms the n-gram approach.
However, their experiment focused solely on file-level defect prediction---not Just-In-Time defect prediction.
Therefore, we wish to investigate if our approach is more effective than the two-phase approach for Just-In-Time defect prediction.

Considering the limitations yet high impact of prior work, we propose \name---a machine learning-based Just-In-Time defect prediction approach that can predict both defect-introducing commits and their associated defective lines.
Then, we formulate the following research questions:
\begin{enumerate}[RQ1)]
\item \rqi
\item \rqii
\item \rqiii
\item \rqiiii
\end{enumerate}

%% file: sections/approach.tex
% Section 5.2 https://arxiv.org/pdf/2009.03612.pdf

\section{\name: A JIT Defect Prediction Approach at the Commit and Line Levels}\label{sec:approach}

In this section, we present the implementation of our \name~approach.
The goal of our \name~approach is to predict defect-introducing commits and identify lines that are associated with that defect-introducing commit (i.e., defective lines).
The underlying intuition of our approach is that code tokens that frequently appeared in defect-introducing commits in the past are likely to be fixed in the future.

\noindent \underline{\textbf{Overview.}} 
Our approach begins with extracting source code tokens of code changes as features (i.e., token features).
Since our JIT defect datasets are highly imbalanced (i.e., 8\%-13\% defective ratio), we apply a SMOTE technique that is optimized by a Differential Evolution (DE) algorithm to handle the class imbalance issue on a training dataset.
Then, we build commit-level JIT defect prediction model using the rebalanced training dataset.
Next, we generate a prediction for each commit in a testing dataset.
After that, we normalize the prediction score by the amount of code changes (i.e., churn) in order to consider the inspection effort when generating the ranking of defect-introducing commits.
For each commit in the testing dataset, we extract the importance score of each token features using a state-of-the-art model-agnostic technique, i.e., Local Interpretable Model-Agnostic Explanations (LIME).
Finally, we rank defective lines that are associated with a given commit based on the LIME's importance scores.
We describe each step in details below.

\textbf{(Step 1) Extracting Bag-of-Tokens Features.} 
Following the underlying intuition of our approach, we represent each commit using Bag-of-Tokens features (i.e., the frequency of each code token in a commit).
To do so, for each commit, we first perform a code tokenization step  to break each changed line into separate tokens.
% Then, we employ the NLTK library to parse its removed lines or added lines into a sequence of tokens.
Then, we parse its removed lines or added lines into a sequence of tokens.
As suggested by Rahman~\ea~\cite{rahman2019natural}, removing these non-alphanumeric characters will ensure that the analyzed code tokens will not be artificially repetitive.
Thus, we apply a set of regular expressions to remove non-alphanumeric characters such as semi-colon (;) and equal sign (=).
We also replace the numeric literal and string literal with a special token (i.e., \texttt{$<$NUM$>$} and \texttt{$<$STR$>$} respectively) to reduce the vocabulary size.
Then, we extract the frequency of code tokens for each commit using the \texttt{Countvectorize} function of the Scikit-Learn Python library.
We neither perform lowercase, stemming, nor lemmatization (i.e., a technique to reduce inflectional forms) on our extracted tokens, since the programming language of our studied systems is case-sensitive.
Otherwise, the meaning of code tokens may be discarded if stemming and lemmatization are applied.

\textbf{(Step 2) Handling class imbalance using an Optimized SMOTE technique.}
Since our JIT defect datasets are highly imbalanced (i.e., 8\%-13\% defective ratio), we apply a SMOTE technique that is optimized by a Differential Evolution (DE) algorithm to handle the class imbalance issue on a training dataset.
The training dataset is splitted into a new training set and a validation set.
The new training set is used to train DE+SMOTE, while the validation set is used to select the best hyper-parameter settings.
We select the SMOTE technique, as prior studies have shown that the SMOTE technique outperforms other class rebalancing techniques~\cite{tantithamthavorn2020impact,agrawal2018better}.
% I think we may remove the explanation of how SMOTE works if the space is not enough
% because we can insert SMOTE citation then let readers read from it :)

The SMOTE technique starts with a set of minority class (i.e., defect-introducing commits). 
For each of the minority class of the training datasets, SMOTE calculates the $k$-nearest neighbors. 
Then, SMOTE selects $N$ instances of the majority class (i.e., clean commits) based on the smallest magnitude of the euclidean distances that are obtained from the k-nearest neighbors. 
Finally, SMOTE combines the synthetic oversampling of the minority defect-introducing commits with the undersampling the majority clean commits.
We use the implementation of \texttt{SMOTE} function provided by the \texttt{Imbalanced-Learn} Python library~\cite{Imblearn}.
% swith a random state of 42.
% and other default parameters setting.
However, prior studies pointed out that the SMOTE technique involves many parameters settings (e.g., $k$ the number of neighbors, $m$ the number of synthetic examples to create, $r$ the power parameter for the Minkowski distance metric), which often impact the accuracy of prediction models~\cite{fu2016tuning,agrawal2018better,tantithamthavorn2020impact,tantithamthavorn2016automated}.

To ensure that we achieve the best performance of the SMOTE algorithm, we optimize the SMOTE technique using a Differential Evolution (DE) algorithm (as suggested by Agrawal~\ea~\cite{agrawal2018better} and  Tantithamthavorn~\ea~\cite{tantithamthavorn2020impact}). 
DE~\cite{Rainer1997} is an evolutionary-based optimization technique, which is based on a differential equation concept. 
Unlike a Genetic Algorithm technique that uses crossover as search mechanisms, a DE technique uses mutation as a search mechanism. 
First, DE generates an initial population of candidate setting of SMOTE's $k$ nearest neighbors with a range value of 1-20.
Then, DE generates new candidates by adding a weighted difference between two population members to the third member based on a crossover probability parameter.
Finally, DE keeps the best candidate SMOTE's parameter setting that is evaluated by a fitness function of maximizing an AUC value for the next generation.
We use the implementation of the differential evolution algorithm provided by Scipy Python library \cite{SciPy}.
As suggested by Agrawal~\ea~\cite{agrawal2018better}, we set the population size to 10, the mutation power to 0.7 and a crossover probability (or \texttt{recombination} parameter in Scipy) to 0.3.

\textbf{(Step 3) Building commit-level JIT defect prediction models.} 
We build a commit-level JIT defect model using both the Bag-of-Tokens features from Step 1 and the commit-level metrics from McIntosh and Kamei~\cite{McIntosh2018}.
The details of commit-level metrics are provided in the replication package.
Prior work found that different classification techniques often produce different performance measures.
Thus, we conduct an experiment on different classification techniques.
We consider the following well-known classification techniques~\cite{tantithamthavorn2016automated, agrawal2018better,  agrawal2019dodge,  tantithamthavorn2018impact}, i.e., Random Forest (RF), Logistic Regression (LR), Support Vector Machine (SVM), k-Nearest Neighbours (kNN), and AdaBoost.
For each project, we build the JIT model using the implementation provided by Python Scikit-Learn package. 
% \cite{scikit-learn}.
We find that LR, kNN, and SVM cannot be built with the Qt project due to the high-dimensional feature space, and the model training time for such models (which takes few hours) is considerably larger than RF (which takes few minutes).
Therefore, we only select the Random Forest classification technique for our study.
After we experiment with different parameter settings of trees (a range of 50 to 1,000), we find that our approach is not sensitive to the parameter setting of random forest.
Thus, we set the number of tress of random forest to 300.
% with a random seed of 42 to mitigate the randomization bias.

% \todo{need to compare with others? , too short here?, combine all or no?}
% \todo{mention that build the model with metrics + show table metrics (but it has same AUC so not sure if this should be included??)}.
% \todo{mention other classifiers then state that RF has the best AUC. (right now KNN and SVM are used, thinking about using MLP, naive bayes, bagging, adaboost)}

% \todo{add more details here}
% The predicted commits are preprocessed similar to during training phase.
% Then they are fed to the trained model to obtain a defective probability.
% The commits are predicted as defective if their probability is greater than 0.5 .

\textbf{(Step 4) Computing a defect density of each commit.}
We then generate the prediction probability for each commit in the testing dataset using the \texttt{predict\_proba} function provided by the Scikit-learn Python library.
Then, we compute the defect density as the probability score normalized by the total changed lines of code of that commit $(\frac{\mathrm{Y}(m)}{\mathrm{\#LOC}(c)})$.
The use of defect density is suggested by prior studies \cite{mende2010effort,Kamei2013} who argued that the cost of applying quality assurance activities may not be the same for each code changes. 
In other words, a prediction model that prioritizes the largest commit as most defect prone would have a very high recall, i.e., those commits likely contain the majority of defects, yet inspecting all those commits would take a considerable amount of time.
In contrast, a model that recommends slightly less defect-prone commits that are smaller to inspect would be more cost-effective~\cite{Kamei2013}.

% , we use the number of changed lines of code as a measure of reviewing effort.

% The commits are then ranked in descending order by the defect density.

% To ensure that we generate an effort-aware ranking of defect-introducing commits, 

% \kla{miss per commit, rewrite- and say compute defect density, clarify that (thee whole test data do not need to be available.)}

% \todo{talk about LIME + Sum + All Tokens}

% We build LIME by using trained random forest classifier and training commits.
% Since the change metrics are not available, only bag-of-word features are used to train LIME.
% In evaluation phase, for a given defective commit, the score of tokens that determine the commit as defective is obtained.
% Then the summation of score in each line is calculated.
% Note that the missing score of some tokens, which is not available in LIME, is treated as zero.
% The score of all lines of the given commit is then used for line-level evaluation.

% find X,Y from Qt and Openstack
\textbf{(Step 5) Generating a ranking of defective lines for a given commit.}
In our studied projects, we found that the average size of the commit varies from 73 to 140 changed lines, but the average ratio of actual defective lines is as low as 51\%-53\%.
Thus, developers still spend unnecessarily effort on locating actual defective lines of that commit~\cite{wattanakriengkrai2020predicting}.
To address this challenge, we propose to generate a ranking of defective lines for a given commit.
For each commit, we compute the importance score of token features using a Local Interpretable Model-agnostic Explanations (LIME) technique.
LIME~\cite{LIME} is a model-agnostic technique that aims to mimic the behavior of the predictions of the defect model by explaining the individual predictions.
Given a commit-level JIT defect prediction model and a commit in the testing dataset, LIME performs the following steps:

\begin{enumerate}
	\item {Generate neighbor instances of a test instance $x$.} LIME randomly generates $n$ synthetic instances surrounding the test instance $x$ using a random perturbation method with an exponential kernel function on cosine distance. 
	\item {Generate labels of the neighbors using a commit-level JIT defect prediction model.} LIME uses the commit-level JIT defect prediction model to generate the predictions of the neighbor instances.
	\item {Generates local explanations from the generated neighbors.}
    LIME builds a local sparse linear regression model (K-Lasso) using the randomly generated instances and their generated predictions from the commit-level defect model.
    The coefficients of the K-Lasso model indicate the importance score of each feature on the prediction of a test instance according to the K-Lasso model.
\end{enumerate}

The LIME's importance score of each token feature ranges from -1 to 1.
A positive LIME score of a token feature ($0<e\leq1$) indicates that the feature has a positive impact on the estimated probability of the test instance (i.e., \textbf{risky tokens}).
On the other hand, a negative LIME score of a token feature ($-1\leq e<0$) indicates that the token feature has a negative impact on the estimated probability (i.e., \textbf{non-risky tokens}).
Once the importance score of each token is computed, we  generate the ranking of defect-prone lines using the summation of the importance score for all tokens that appear in that line.
We use the implementation of LIME provided by the \texttt{lime} Python package.

%% file: sections/experiment.tex
\begin{table}[t]
\caption{The statistics of our studied datasets.}
\label{tab:dataset}
\resizebox{\columnwidth}{!}{%
\begin{tabular}{l|c|c|c|c|c|}
\cline{2-6}
\multicolumn{1}{c|}{}                    & \textbf{\#Commits} & \textbf{\begin{tabular}[c]{@{}c@{}}\%Defect-\\ Introducing \\ Commits\end{tabular}} & \textbf{\begin{tabular}[c]{@{}c@{}}\#Unique \\ Tokens\end{tabular}} & \textbf{\begin{tabular}[c]{@{}c@{}}Avg. of \\Commit \\ Size\end{tabular}} & \textbf{\begin{tabular}[c]{@{}c@{}}Avg. of \\ \%Defective \\ Lines\end{tabular}} \\ \hline
\multicolumn{1}{|l|}{\textbf{Openstack}} & 12,374              & 13\%                                                                                & 32K                                                            & 73 LOC                                                         & 53\%                                                                 \\ \hline
\multicolumn{1}{|l|}{\textbf{Qt}}        & 25,150              & 8\%                                                                                  & 81K                                                           & 140 LOC                                                       & 51\%                                                                   \\ \hline
\end{tabular}%
}
\end{table}

% \subsection{Parameter Setting}

% Our model implementation is based on various Python libraries \cite{scikit-learn, NumPy, SciPy}.
% We extract unigram BoW feature from code change by using CountVectorizer in Scikit-Learn, with min\_df = 1.
% We use SMOTE, implemented in Imbalanced-Learn \cite{Imblearn}, to generate additional defective commits samples during training phase.
% We apply differential evolution algorithm in Scipy to find the best number of neighbors of SMOTE.
% The range of this value is [1,20], similar to the work in \cite{Agrawal2018}
% However, there are two limitations of the algorithm in Scipy.
% First of all, it does not support finding optimal integer value for an objective function.
% Secondly, it can only find the best solution for minimization tasks.
% To solve the first issue, the value obtained from the algorithm is rounded to nearest integer.
% We overcome the second issue by having the objective function returns negative AUC, which can be expected the same positive value when training final model.
% At the end, the model of our choice is default RandomForest classifier, which can be found in Scikit-Learn.
% We use the same train/test data proportion as in \cite{CC2Vec}.

% \subsection{Evaluation Measures}

% We divide evaluation metrics into two categories: traditional metrics and effort-aware metrics. 
% The first category evaluates performance of the model in different aspects, while the latter measures the ability of the model to reduce effort of software engineer and developers while looking for a defective part. 

%% file: sections/results.tex
%%%%%%%%%%%%%%%%%%%%%%%%%%%%%%%%
\section{Experimental Setup and Results}\label{sec:results}
%%%%%%%%%%%%%%%%%%%%%%%%%%%%%%%%

In this section, we describe the studied datasets and present the experimental results with respect to our four research questions.

\textbf{Studied Datasets.}
In this paper, we select the dataset of McIntosh and Kamei~\cite{McIntosh2018} due to the following reasons.
First, we would like to establish a fair comparison, using the same training and testing datasets with previous work~\cite{hoang2019deepjit,CC2Vec}, where this dataset was used.
% With this dataset, we are able to use the same training and testing datasets as the previous work~\cite{hoang2019deepjit,CC2Vec}.
Second, we would like to ensure that our results rely on high quality datasets. 
Recently, researchers raised concerns that the SZZ algorithm~\cite{SZZalgo} may produce many false positives and false negatives~\cite{rodriguez2018reproducibility}.
However, the datasets of McIntosh and Kamei~\cite{McIntosh2018} have been manually verified through many filtering steps (e.g., ignore comment updates, ignore white space/indentation changes, remove mislabelled defect-introducing commits).
% \kla{talk about noisy SZZ algorithms then McIntosh Kamei has conduct a manual verification so we can ...trust :) the datasets}
% The datasets are originally collected by McIntosh and Kamei, using SZZ algorithm \cite{SZZalgo}. 
% The SZZ algorithm takes the following steps to obtain defective lines of defect-introducing commits.
% First, SZZ algorithm identifies defect-fixing commits from the given defect-introducing commits.
% Then the algorithm use \texttt{diff} command to pinpoint the modified lines of defect-fixing commits.
% After that the \texttt{blame} command is used to detect the version history that introduces the modified lines.
% It is possible that some code changes in the datasets, collected by using SZZ algorithm \cite{SZZalgo}, are mislabelled.
% Previous work \cite{rodriguez2018reproducibility} found that SZZ algorithm may classify clean commits as defect-introducing commits by observing code style changes (i.e., white space, indentation, comments, etc.) or missing authorship information.
% In addition, defect-introducing commits may be linked to defect report, which does not report actual defects.
% Hence, to prevent false positive (i.e., clean commits being identified as defective), McIntosh and Kamei \cite{McIntosh2018} analyzed the datasets then filtered out some mislabelled defect-introducing commits.
% \kla{write below, why you choose the McIntosh and Kamei, put justification, how good is this dataset? read the McIntosh/Kamei paper about the preliminary analysis and Table 3}
Finally, we select the datasets of McIntosh and Kamei~\cite{McIntosh2018} with two open-source software systems, i.e., OpenStack and Qt.
Openstack is an opensource software for cloud infrastructure service.
Qt is a cross-platform application development framework written in C++. 
% It supports software development for desktop and mobile.
Table~\ref{tab:dataset} presents the statistics of the studied datasets.

% We use 12,374 commits of Openstack from 11/2011 to 02/2014, and 25,150 commits of Qt from 06/2011 to 03/2014, to conduct experiment of our study. 

Below, we present the approach and the results with respect to our four research questions.

\input{sections/RQ1.tex}
\input{sections/RQ2.tex}
\input{sections/RQ3.tex}

\input{sections/RQ4.tex}

%% file: sections/RQ1.tex
%%%%%%%%%%%%%%%%%%%%%%%%%%%%%%%%
\subsection*{\textbf{(RQ1) \rqi}}
%%%%%%%%%%%%%%%%%%%%%%%%%%%%%%%%

% incldue this table in research study table
% then remove this table
% \begin{table*}
%   \centering
%   \caption{RQ2 result}
%     \begin{tabular}{l|c|c|c|c|c|c|c|c|c|c|c|c|}
%     \cline{2-13}    
%      & \multicolumn{6}{c|}{Openstack} & \multicolumn{6}{c|}{QT} \\
%     \cline{2-13}    
%      & Precision & Recall & F1 & AUC & FAR & d2h & Precision & Recall & F1 & AUC & FAR & d2h \\
%     \hline
%     Our approach & 0.48 & 0.33 & 0.39 & 0.83 & 0.05 & 0.47 & 0.43 & 0.16 & 0.23 & 0.82 & 0.02 & 0.60 \\
%     LR    & 0.17 & 0.01 & 0.01 & 0.48 & 0.00 & 0.70 & 0.57 & 0.02 & 0.04 & 0.64 & 0.00 & 0.69\\
%     % RF (metrics) & 0.44 & 0.15 & 0.22 & 0.82 & 0.53 & 0.10 & 0.17 & 0.81 \\
%     DeepJIT & 0     & 0     & 0     & 0.75  & 0 & 0.71 & 0.35  & 0.03  & 0.06  & 0.76 & 0.01 & 0.69\\
%     CC2vec [Train only] & 0.14  & 0.99  & 0.24  & 0.77  & 0.87 & 0.61 & 0.1 & 0.96  & 0.19  & 0.81 & 0.63	& 0.45\\
%     CC2Vec [Train+Test] & 0.27 & 0.70 & 0.39 & 0.80 & 0.26 & 0.28 & 0.24 & 0.70 & 0.35 & 0.84 & 0.17 & 0.25 \\
%     \hline
%     \end{tabular}%
%   \label{tab:RQ1}%
% \end{table*}%

% font in subfigure looks different from the font in paper
\begin{figure*}[t]
    \begin{subfigure}{.5\textwidth}
        \centering
        \includegraphics[width=\columnwidth]{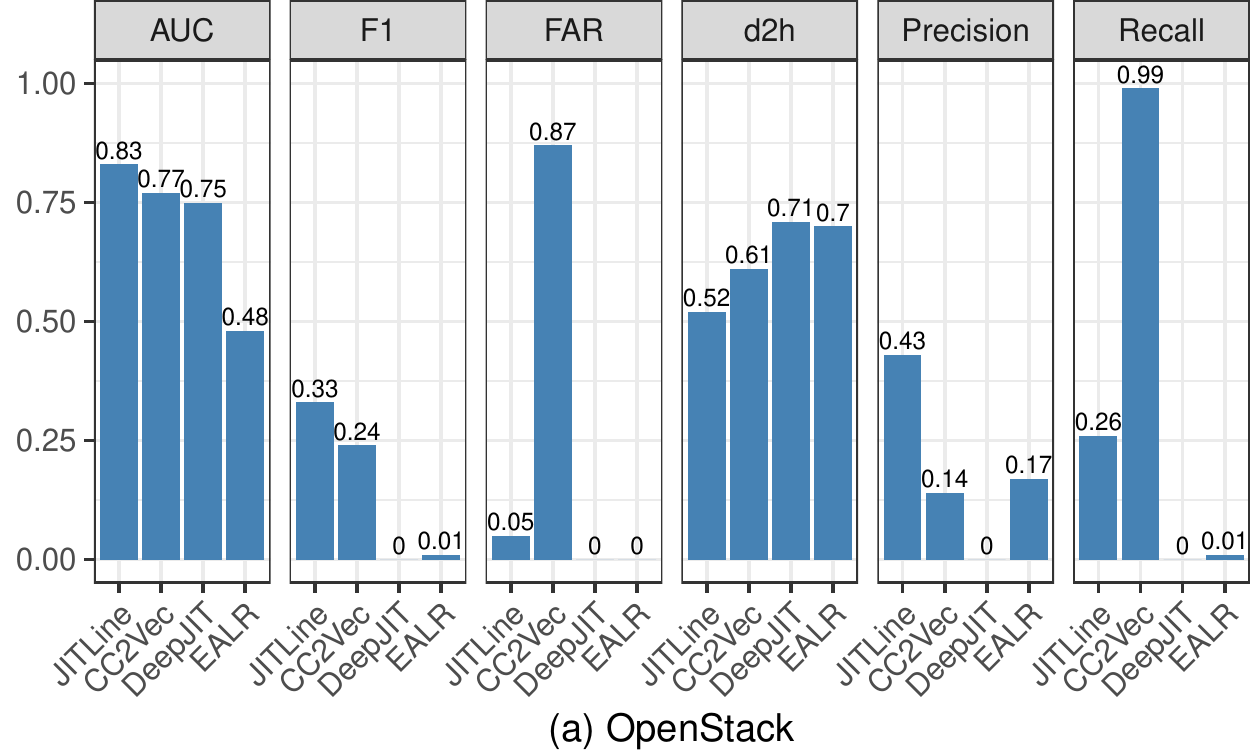}
    \end{subfigure}    
    \begin{subfigure}{.5\textwidth}
        \centering
        \includegraphics[width=\columnwidth]{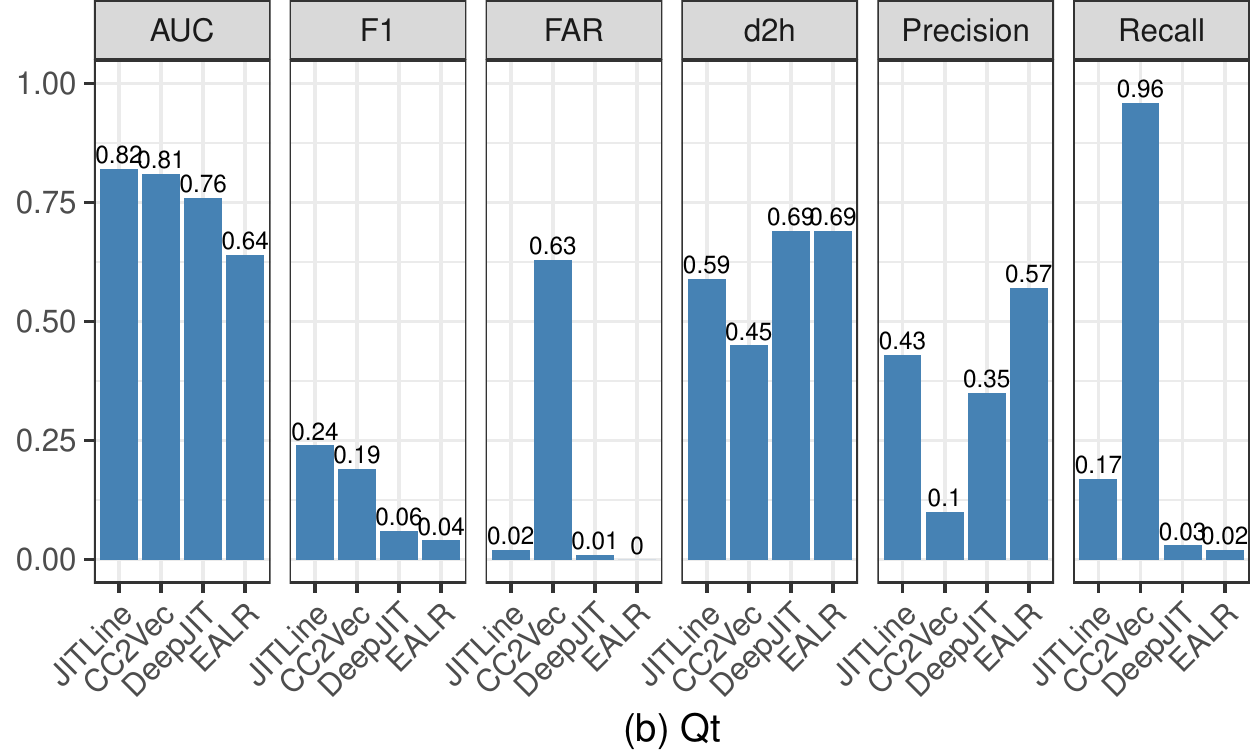}
    \end{subfigure}
    % \hfill 
\caption{(RQ1) The evaluation result of our \name~approach compared with the state-of-the-art approaches for Just-In-Time defect prediction (i.e., CC2Vec(Train only), DeepJIT, and EALR).} %\todo{caption?}}
\label{fig:RQ1_result}
\end{figure*}

% We train random forest model by using code change and commit metrics as features.
% Before training phase, over-sampling technique called SMOTE is performed on training data to increase defective commit samples.
% Differential evolution algorithm is employed to select the best number of neighbors of SMOTE to achieve the highest AUC.

% We use the same data proportion to train and evaluate baselines, but different features are used.
% In particular, we use only commit metrics and code change to train logistic regression classifier and random forest, respectively.
% For CC2Vec and DeepJIT, we rerun the source code to train model with only train data to measure their actual performance.

% During evaluation phase, we calculate precision, recall, F1, FAR and D2H at defective commits probability threshold of 0.50.

% \subsubsection{Classification Metrics}

% This metric category includes precision, recall, F1, AUC.

\smallsection{Approach}
To answer this RQ, we evaluate our \name~using the same training/testing datasets as prior studies~\cite{hoang2019deepjit, McIntosh2018, CC2Vec} to establish a fair comparison.
For training, we use 11,043 commits for OpenStack and 22,579 commits for Qt.
For testing, we use 1,331 commits for OpenStack and 2,571 for Qt.
Since our JIT defect datasets are time-wise, we do not perform cross-validation to avoid the use of testing data in the training data~\cite{jimenez2019importance}.
% In our experiment, we do not train model by using cross-validation scheme because it does not conform real-world scenario.
% To be specific, training ML model with cross-validation scheme may use future code change to train model, which is not available when deploying in real world.
Then, we compare our \name~with the following three state-of-the-art JIT defect prediction approaches (i.e., EARL, DeepJIT, CC2Vec).
The details of the state-of-the-art approach is provided in Section~\ref{sec:background}.
Finally, we evaluate these approaches using the following six traditional evaluation measures~\cite{hoang2019deepjit, wattanakriengkrai2020predicting, CC2Vec, agrawal2018better}.

% put the outstanding measure at the top of the list
\begin{enumerate}

\item AUC is an Area Under the ROC Curve (i.e., the true positive rate and the false positive rate). AUC values range from 0 to 1, with a value of 1 indicates perfect discrimination, while a value of 0.5 indicates random guessing.

\item F-measure is a harmonic mean of precision and recall, which is computed as $ \frac{2\times\mathrm{Precision}\times\mathrm{Recall}}{\mathrm{Precision}+\mathrm{Recall}}$. 
We use the probability threshold of 0.5 for calculating precision and recall.

\item False Alarm Rate (FAR)~\cite{agrawal2019dodge} measures the ratio of incorrectly predicted defect-introducing commits and the number of actual clean commits $\frac{\mathrm{FP}}{\mathrm{FP}+\mathrm{TN}}$. The lower the FAR value is, the fewer the incorrectly predicted defect-introducing commits that developers need to review. 
In other words, a low FAR value indicates that developers will spend
less effort on reviewing the incorrectly predicted defect-introducing commits.
% FAR indicates the ratio of incorrectly predicted commits as defectives against all prediction. This measure can be calculated as $FAR = \frac{\text{\# incorrectly predicted commits as defectives}}{\text{\# defective commits}}$. The higher value, the more possibility that clean commits are predicted as defective commits.

\item Distance-to-Heaven (d2h)~\cite{agrawal2019dodge} is a root mean square of recall and FAR values, which can be computed as $\sqrt{\frac{(1-\mathrm{Recall})^2 + (0-\mathrm{FAR})^2}{2}}$. 
A d2h value of 0 indicates that an approach can correctly predict all defect-introducing commits without any false positive.
A high d2h value indicates that an approach is far from perfect, e.g., achieving a high recall value but also have high FAR value and vice versa.

\item Precision measures the ability of an approach to correctly predict defect-introducing commits, which can be calculated as follows: $\mathrm{Precision} = \frac{\mathrm{TP}}{\mathrm{TP+FP}}$. The higher precision, the better model to correctly predict defect-introducing commits. 

\item Recall measures the ability to correctly retrieve defect-introducing commits when making a prediction. The calculation of this measure is $\mathrm{Recall} = \frac{\mathrm{TP}}{\mathrm{TP+FN}}$. High recall indicates that the model can obtain a lot of defect-introducing commits during prediction. 
\end{enumerate}

% After the training phase finished, we compare precision, recall, F1 and AUC of our approach and baselines.

% \todo{put openstack result first then followed by Qt (do this with the whole paper...}

% not sure about percentage change from 0.00 to 0.36, is it 3600% change?
\smallsection{Results} 
\textbf{Our \name~approach achieves an AUC 28\%-73\% higher and an F-measure 26\%-38\% higher than the state-of-the-art approaches (i.e., CC2Vec).}
% written by Oat
% \textbf{Our \name~approach achieves AUC 28\%-73\% higher than the state-of-the-art approaches.}
Figure~\ref{fig:RQ1_result} presents the experimental results of our approach and the state-of-the-art approaches with respect to six evaluation measures, i.e., AUC, FAR, d2h, precision, and recall for Openstack and Qt.
% This figure reports precision, recall, F1, AUC, FAR and D2H of Qt and Openstack, obtained from our approach and baselines. 
We find that our \name~approach achieves the highest AUC value of 0.83 for Openstack and 0.82 for Qt, which is 1\%-8\% higher than CC2Vec, 8\%-10\% higher than DeepJIT, and 28\%-73\% higher than EALR.
We also find that our \name~approach achieves the highest F-measure value of 0.33 for Openstack and 0.24 for Qt, which are 26\%-38\% higher than CC2Vec, 300\%-3,300\% higher than DeepJIT, and 500\%-3,200\% higher than EALR.
% From the evaluation result of Qt in Fig \ref{fig:RQ1_result} (a), our approach achieves the highest F1 of 0.23 and highest AUC of 0.82 when compared to other approaches.
% The AUC of CC2Vec and DeepJIT, are 1\% and 9\% lower than JITLine's AUC.
% Furthermore, our approach obtains 21\% and 283\% higher F1 of Qt and Openstack, respectively. 
This finding indicates that our approach outperforms the state-of-the-art approaches in terms of AUC and F-measure.
% ability to discriminate defective commits of our approach and deep learning-based approaches are comparable.

\textbf{Our \name~approach achieves a False Alarm Rate (FAR) 94\%-97\% lower than the CC2Vec approach.}
We find that our \name~approach achieves a False Alarm Rate (FAR) of 0.05 for Openstack and 0.02 for Qt, which is similar to DeepJIT (FAR=0.01) and EALR (FAR=0).
Similarly, our \name~approach also achieves a d2h of 0.52 for OpenStack and 0.59 for Qt, which is lower than the state-of-the-art approaches (i.e., DeepJIT and EALR).
However, we observe that the d2h of our approach for Qt project is higher than the CC2Vec approach.
For Qt project, we find that the lower d2h value of the CC2Vec approach has to do with the high recall value of 0.96---i.e., the CC2Vec approach predicts most of the commits as defect-introducing, but 63\%-87\% of them are incorrect (i.e., many of them are false positives)--- indicating that developers may spend unnecessarily effort to inspect actual clean commits that are incorrectly predicted as defect-introducing commits when the CC2Vec approach was used.
On the other hand, the high d2h value of our approach has to do with the low recall of 0.16, but our approach achieves a low FAR of 0.02, indicating that the predictions from our \name~approach is less likely to predict actual clean commits as defect-introducing.
After considering both the ability of identifying defect-introducing commits (i.e., Recall) and the additional costs (i.e., FAR), our approach still outperforms state-of-the-art approaches (i.e., CC2Vec (only for OpenStack), DeepJIT, and EALR). 

%% file: sections/RQ2.tex
\subsection*{\textbf{(RQ2) \rqii}}

\smallsection{Approach}
To answer this RQ, we evaluate our \name~ and compare with the four state-of-the-art JIT defect prediction approaches (as mentioned in RQ1) using the following cost-effective measures~\cite{mende2010effort,Kamei2013,agrawal2019dodge,huang2017supervised,yang2016effort}:

\begin{enumerate}
\item PCI@20\%LOC measures the proportion of actual defect-introducing commits that can be found given a fixed amount of effort, i.e., the Top 20\% LOC of the whole project.
A high value of PCI@20\%LOC indicates that an approach can rank many actual defect-introducing commits so developers will spend less effort to find actual defect-introducing commits.
% how a model can obtain defective lines in defective commits when reading 20\% cummulative LOC of the whole project. The higher the value is, the more defective commits obtained by spending effort to observe commits of which number of LOC is within 20\% of cumulative LOC of all release in a project. This measure is calculated as $Recall@20\%Effort = \frac{\text{\# defective commits within 20\% cummulative LOC}}{\text{\# total defective commits}}$.

\item Effort@20\%Recall measures the amount of effort (measured as LOC) that developers have to spend to find the actual 20\% defect-introducing commits divided by the total changed LOC of the whole testing dataset.
A low value of Effort@20\%Recall indicates that the developers will spend a little amount of effort to find the 20\% actual defect-introducing commits.

\begin{figure}[t]
    % \hfill 
    \begin{subfigure}{\columnwidth}
        \centering
        \includegraphics[width=\columnwidth]{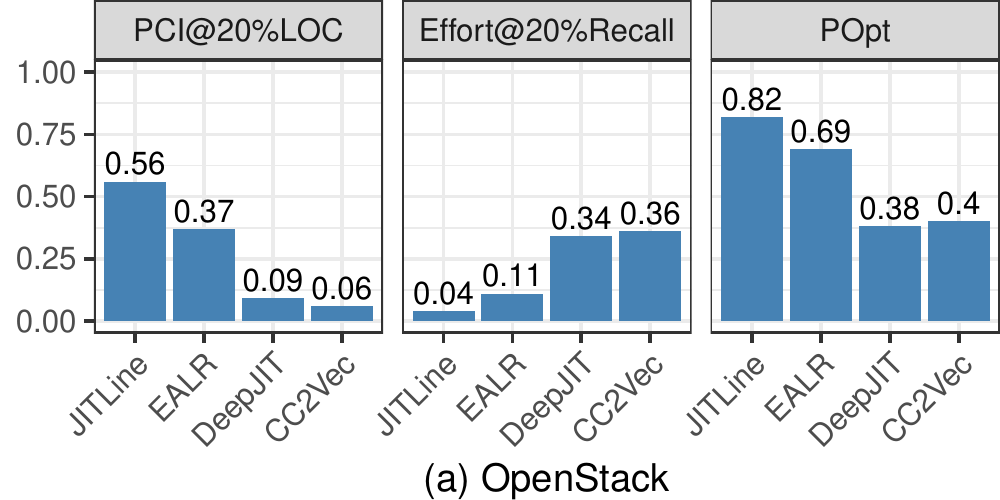}
        % \caption{openstack result}
    \end{subfigure}
    \begin{subfigure}{\columnwidth}
        \centering
        \includegraphics[width=\columnwidth]{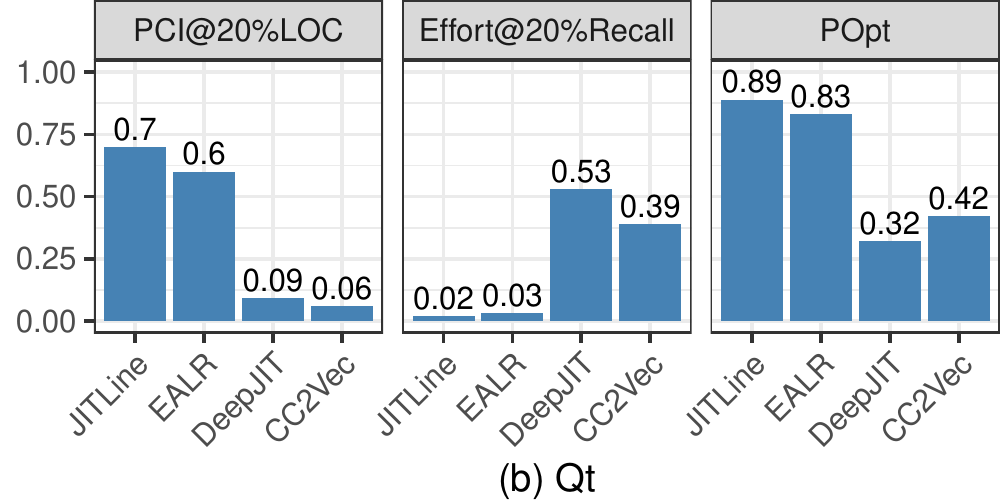}
        % \caption{qt result}
    \end{subfigure}
\caption{(RQ2) The cost-effectiveness of our \name~approach compared to the state-of-the-art approaches for Just-In-Time defect prediction with respect to PCI@20\%Recall, Effort@20\%Recall, and P\textsubscript{Opt}.}
\label{fig:rq2}
\end{figure}

\item  P\textsubscript{opt} is defined as 1-$\Delta_{\mathrm{opt}}$, where $\Delta_{\mathrm{opt}}$ is the area of the effort-based (i.e., churn) cumulative lift chart between an optimal model and a prediction model. 
The effort-based (i.e., churn) cumulative lift chart is the relationship between the cumulative percentage of defect-introducing commits from a prediction model ($y$-axis) and the cumulative percentage of the inspection effort ($x$-axis).
Similar to prior studies~\cite{mende2010effort,agrawal2019dodge,yang2016effort}, we use the normalized version of P\textsubscript{opt}, which is defined as $1-\frac{\mathrm{Area}(\mathrm{Optimal})-\mathrm{Area}(\mathrm{Our})}{\mathrm{Area}(\mathrm{Optimal})-\mathrm{Area}(\mathrm{Worst})}$.
For the \emph{optimal} model and the \emph{worst} model, all commits are ranked by the actual defect density in descending and ascending order, respectively.
For \emph{our} model, all commits are ranked by the estimated defect density $(\frac{\mathrm{Y}(m)}{\mathrm{\#LOC}(c)})$ in descending order.
% https://scikit-learn.org/stable/modules/generated/sklearn.metrics.auc.html
\end{enumerate}

\smallsection{Results}
\textbf{Our \name~approach is 17\%-51\% more cost-effective than the state-of-the-art approaches in term of PCI@20\%LOC.} 
% The experiment result summary is reported in Fig \ref{fig:rq2}.
Figure~\ref{fig:rq2} presents the cost-effectiveness of our \name~approach compared to the state-of-the-art approaches for Just-In-Time defect prediction with respect to the PCI@20\%LOC, Effort@20\%Recall and P\textsubscript{opt} measures.
We find that our \name~approach is more cost-effective than the state-of-the-art approaches for three cost-effectiveness measures.
We find that our \name~achieves a PCI@20\%LOC of 0.56 for Openstack and 0.70 for Qt, while the state-of-the-art achieves a PCI@20\%LOC of 0.06-0.37 for OpenStack and 0.06-0.60 for Qt.
This finding indicates that given a fixed amount of inspection effort at 20\%LOC, our \name~approach can correctly predict 17\%-51\% higher number of actual defect-introducing commits than the state-of-the-art approaches.

\textbf{Our \name~approach can save the amount of effort by 89\%-96\% to find the same number of actual defect-introducing commits (i.e., 20\% Recall) when compared to the state-of-the-art approaches.}
Our \name~approach achieves an Effort@20\%Recall of 0.04 for Openstack and and 0.02 Qt, while the state-of-the-art approaches achieve an Effort@20\%Recall of 0.11-0.36 for Openstack, and 0.03-0.53 for Qt.
% According to the result, it is implied that our approach is the best to reduce effort when inspecting top 20\% defective lines of all bug-introducing commits, when compared to the other baselines.
% \todo{explain results as above}
Similarly, our \name~approach achieves a P\textsubscript{opt} of 0.82 for OpenStack and 0.89 for Qt, which is 116\% and 178\% higher than the state-of-the-art approaches for OpenStack and Qt, respectively.
% \kla{Oat will do this?}
In particular, our P\textsubscript{opt} is 7\% to 19\% higher than EALR, 116\% to 178\% higher than DeepJIT, and 105\% to 112\% higher than CC2Vec.
This finding suggests that, to find the same amount of actual defect-introducing commits, our \name~approach can reduced the amount of effort by 85\% and 96\% when compared to the state-of-the-art approaches, which may provide the best return on investment.

%% file: sections/RQ3.tex
\subsection*{\textbf{(RQ3) \rqiii}}
%%%%%%%%%%%%%%%%%%%%%%%%%%%%%%%%

\begin{table}[t]
\caption{(RQ3) The average CPU and GPU computational time (minutes$\pm$95\% Confidence Interval) of the model training of JIT defect prediction approaches after repeating the experiment 5 times.}
\label{tab:RQ3}
\resizebox{\columnwidth}{!}{ 
  \centering
\begin{tabular}{l|c|c|c|c|}
\cline{2-5}
    & \multicolumn{2}{c|}{\textbf{CPU}} & \multicolumn{2}{c|}{\textbf{GPU}}         \\
    \cline{2-5} 
    & \textbf{Openstack}  & \textbf{Qt} & \textbf{Openstack} & \textbf{Qt} \\ 
    \hline
\multicolumn{1}{|l|}{\textbf{\name}} & 36$\pm$1 secs             & 175$\pm$1 secs       & -                  & -           \\
\multicolumn{1}{|l|}{\textbf{DeepJIT}}              & 70$\pm$7 mins            & 143$\pm$7 mins    & 2$\pm$0.01 mins  & 5$\pm$0.01 mins     \\
\multicolumn{1}{|l|}{\textbf{CC2Vec}}               & 146$\pm$16 mins             & 300$\pm$6 mins  & 13$\pm$0.05 mins   & 30$\pm$0.10 mins \\ \multicolumn{1}{|l|}{\textbf{EARL}}               & 8$\pm$1 secs             & 97$\pm$1 secs  & -   & -  \\
\hline
\end{tabular}
}
\end{table}

% The process in this RQ is similar to the steps done in RQ1.
% However, we only measure training time of CC2Vec and DeepJIT when running on CPU and GPU.
% Specifically, since DeepJIT is integrated with CC2Vec, we measure both training time of CC2Vec and DeepJIT.
% Then both time stamp are combined as final training time of CC2Vec.

\smallsection{Approach} To answer this RQ, we measure the CPU computational time of the model training of our approach, and the CPU and GPU computational time of the model training of deep learning approaches (i.e., DeepJIT and CC2Vec).
For our approach, we set \texttt{n\_jobs} argument of the \texttt{RandomForestClassifier} function of Scikit-Learn library to -1 to ensure that all available CPU cores are used in parallel.
For the deep learning baselines, we use \texttt{cpu} function provided by the Pytorch deep learning library to ensure that all available CPU cores are used in parallel. 
% \todo{I check in htop that cpu runs in parallel, but not sure if I can say this in paper...} / fine with that / .
We perform the experiment using the following equipment: AMD Ryzen 9 5950X 16 Cores/32 Threads Processor, RAM 64GB, NVIDIA GeForce RTX 2080 Ti 11GB.
To ensure that our measurement is accurate and strictly controlled, we reserve the computing resources and ensure that the resources are idle with no other running tasks.
To combat the randomization bias, we repeat the experiment 5 times.

\smallsection{Results} \textbf{Our \name~approach is 70-100 times faster than the deep learning approaches for Just-In-Time defect prediction.}
Table~\ref{tab:RQ3} presents the average CPU and GPU computational time (minutes) of the model training of JIT defect prediction approaches after repeating the experiment 5 times.
We find that the model training time of our \name~approach takes approximately 1-3 minutes, while the model training time of the deep learning approaches for Just-In-Time defect prediction require 1-5 hours (70 to 300 minutes).
Given the same running cost (on CPU), this finding suggests that our approach is more cost-efficient than the deep learning approaches.

The computation time of the deep learning approaches can be accelerated by using a high-end GPU hardware.
However, we find that the model training time of the deep learning approaches on the GPU device is relatively faster than using the CPU hardware with an additional GPU cost.
Nevertheless, the model training time of deep learning approaches on GPU (2-30 minutes) still takes relatively longer than the model training time of our approach on CPU (1-3 minutes).

% When considering the cost-benefits, we find that the benefits of our approach still outweigh the cost.
% \todo{thinking how to end this one?}
% \todo{say something like despite using the GPU, the computational time still x\% more than our approach that use CPU}.
% We found that our approach is much faster than both CC2Vec and DeepJIT when training on GPU and CPU.
% The overall result can be seen in table 

% Time spent to train models on Qt is much higher than training on Openstack. Both DeepJIT and CC2Vec spent about 11 minutes and 24 minutes respectively. However, the CPU time of both approach is about 2 to 6 hours. In contrast, our approach has training time around 1 minute to train model.

% In Openstack, the training time of DeepJIT and CC2Vec, when training on GPU, is about 5 minutes and 13 minutes, respectively. When training on CPU, both DeepJIT and CC2Vec spend more than an hour to train model. However, our approach, trained on 32-core CPU, achieves training time less then one minute. 

% Do I need to conclude here??

%% file: sections/RQ4.tex
%%%%%%%%%%%%%%%%%%%%%%%%%%%%%%%%
\subsection*{\textbf{(RQ4) \rqiiii}}

\begin{figure}[t]
\centering
\includegraphics[width=\columnwidth]{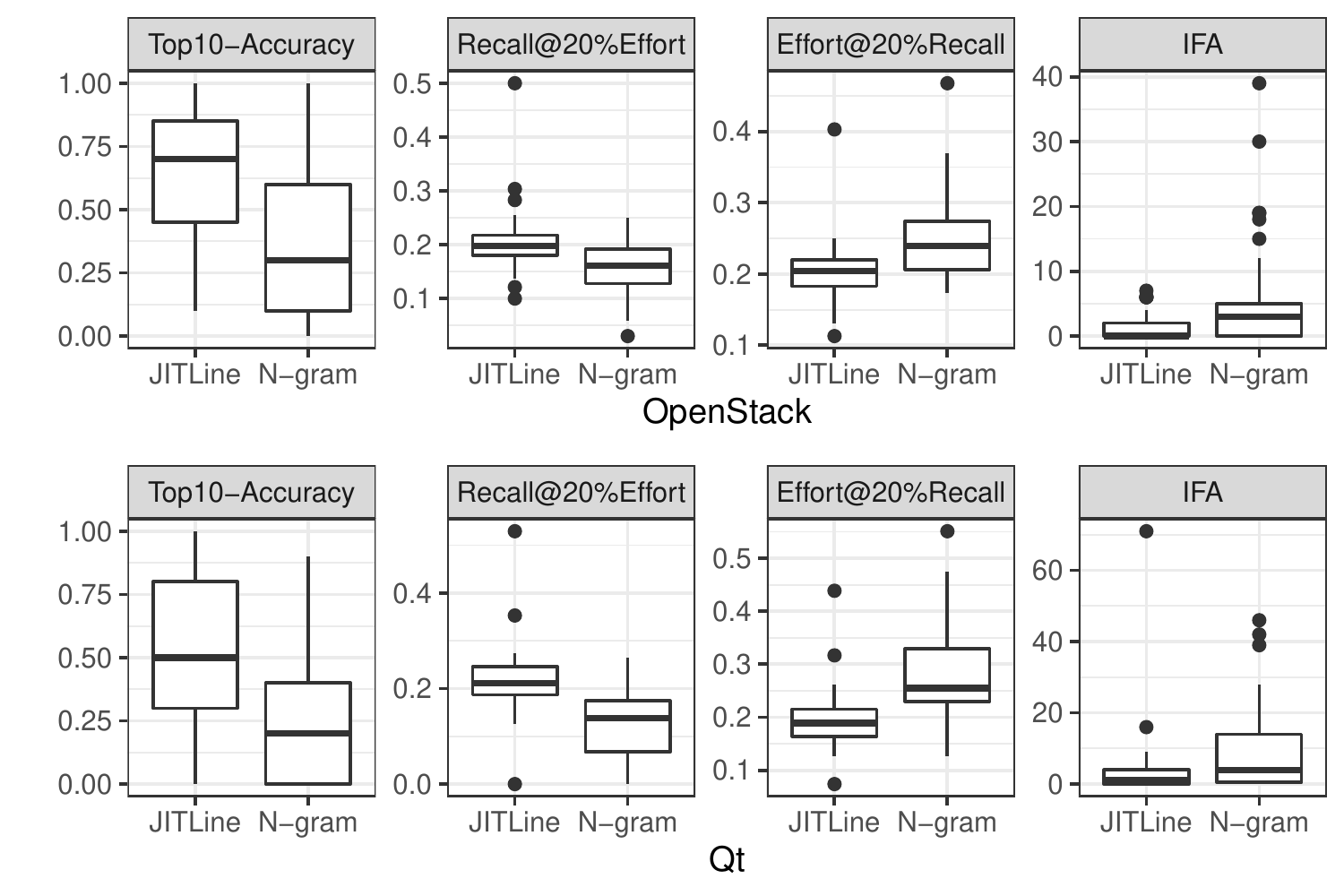}
\caption{(RQ4) The results of our \name~at the line level when compared to the N-gram-based line-level JIT defect prediction approach of Yan~\ea~\cite{yan2020just} with respect to Top-10 Accuracy($\nearrow$), Recall@20\%Effort($\nearrow$), Effort@20\%Recall($\searrow$), and  IFA($\searrow$). The higher ($\nearrow$) or the lower ($\searrow$) the values are, the better the approach is.}
\label{fig:rq4}
\end{figure}

\smallsection{Approach}
To address this RQ, we first need to collect the line-level ground-truth data.
To do so, we start from cloning a git repository of the studied projects.
% \footnote{We used the \texttt{get\_commits\_last\_modified\_lines} function provided by the PyDriller library.}---
Then, we use PyDriller~\cite{PyDriller}, a Python library for mining GitHub repository, to identify defect-fixing commits that are associated with each defect-introducing commit that is provided by McIntosh and Kamei~\cite{McIntosh2018}.
Once identified, we examine the diff (a.k.a. code changes) made by the defect-fixing commits to identify lines that are modified/deleted by defect-fixing commits.
Similar to prior work~\cite{da2016framework,rodriguez2018reproducibility}, the lines that were modified or deleted by defect-fixing commits are identified as defective lines, otherwise clean. 
% We also conduct a manual examination to ensure that (1) \kla{how do you ensure that they are correct, no false positives and false negatives?}.
% We also conduct a manual examination on random defect-introducing commits by checking whether their code change appears on their corresponding defect-fixing commit.
Then, we compare our \name~with the state-of-the-art line-level JIT defect prediction approach by Yan~\ea~\cite{yan2020just}.
We implement the N-gram approach using the implementation provided by Hellendoorn~\ea~\cite{hellendoorn2017deep},
Since Yan~\ea~\cite{yan2020just} found that the Jelinek-Mercer (JM) smoothing method is the best choice, and the N-gram length has no substantial impact on the average performance, we followed their advice by using the Jelinek-Mercer (JM) smoothing method and the N-gram length of 6.
% We apply N-gram model with the same bug-introducing commits that are found by our approach.
% To implement N-gram, instead of using Java lexer to read code changes, we utilize function of \texttt{TokenizedLexer.java}.
% We first train the N-gram model with all code change in training dataset.
% Then the trained model is used to extract line score from bug-introducing commits for evaluation.
% train n-gram using all source code from training data
% then evaluate the test source code with identified buggy commits.
Finally, we evaluate our approach and Yan~\ea~\cite{yan2020just} using the following evaluation measures at the line level~\cite{yan2020just,wattanakriengkrai2020predicting}:

\begin{enumerate}
\item Top-10 Accuracy measures the proportion of actual defective lines that are ranked in the top-10 ranking. 
Traditionally, developers may need to inspect all changed lines for a given commit---which is not ideal when SQA resources are limited.
A high top-10 accuracy indicates that many of the defective lines are ranked at the top, which is considered effective.
% Thus, we wish to evaluate the effectiveness of our line-level recommendations of our approach when compared to the state-of-the-art
% Considering a limited inspection budget, we focus on the Top-10 defective lines.
\item Recall@20\%LOC measures the proportion of defective lines that can be found (i.e., correctly predicted) given a fixed amount of effort (i.e., the top 20\% of changed lines of a given defect-introducing commit).
A high value of Recall@20\%LOC indicates that an approach can rank many actual defective lines at the top.

\item Effort@20\%Recall\textsubscript{line} measures the percentage of the amount of effort that developers have to spend to find the actual 20\% defective lines of a given defect-introducing commit.
A low value of Effort@20\%Recall\textsubscript{line} indicates that the developers will spend a little amount of effort to find the 20\% actual defective lines. 

\item Initial False Alarm (IFA) measures the number of clean lines that developers need to inspect until finding the first actual defective line for a given commit. 
A low IFA value indicates that developers only spend time inspecting only a few number of clean lines to find the first actual defective line.
% , while a high IFA value indicates that developers have to spend unnecessary effort on inspecting clean lines. 
% The intuition behinds this measure is that developers may stop inspecting if they could not get promising results (i.e., find defective lines) within the first few inspected lines~\cite{wattanakriengkrai2020predicting,yan2020just}.
\end{enumerate}

% Thus, many actual defective lines can be found given a fixed amount of effort. 
% On the other hand, a low value of Recall@20\%LOC$_\mathrm{COMMIT}$ indicates many clean lines are in the top20\% LOC.
% Thus, developers need to spend more effort to identify defective lines.
% A high Recall@20\%LOC$_\mathrm{COMMIT}$ indicates that many defective lines are ranked at the top , 

% The higher the value is, the more defective commits obtained by spending effort to observe commits of which number of LOC is within 20\% of cumulative LOC of all release in a project. This measure is calculated as $Recall@20\%Effort = \frac{\text{\# defective commits within 20\% cummulative LOC}}{\text{\# total defective commits}}$.
% In contrast, a high value of Effort@20\%Recall indicates that developers have to spend more effort to find the 20\% actual defective lines. 

\smallsection{Results}
\textbf{Our \name~approach is 133\%-150\% more accurate than the baseline approach by Yan~\ea~\cite{yan2020just} for identifying actual defective lines in the top-10 recommendations.}
Figure~\ref{fig:rq4} shows that our approach achieves a median Top-10 Accuracy of 0.7 for  OpenStack and 0.5 for Qt, while the baseline approach achieves a Top-10 Accuracy of 0.3 for Openstack and 0.2 for Qt.
In addition, we find that our \name~approach can find actual defective lines 25\%-50\% higher than the baseline approach, given the same amount of effort at 20\%LOC.
Figure~\ref{fig:rq4} shows that our approach achieves a median Recall@20\%LOC of 0.20 for OpenStack and 0.21 for Qt, while the baseline approach achieves a median Recall@20\%LOC of 0.16 for OpenStack and 0.14 for Qt.

% done later
Our Wilcoxon signed-ranked test also confirms that the difference of Top-10 Accuracy and Recall@20\%Effort between our approach and the baseline is statistically significantly ($p$-value $<$ 0.05) with a Cliff's $|\delta|$ effect size of large ($|\delta|=0.49-0.67$) for both Top-10 Accuracy and Recall@20\%LOC.
% The statistical test confirms that our \name~approach can find more actual defective lines than the baseline approach given the same amount of inspection effort at 20\%LOC of each commit.

\textbf{Our \name~approach requires 17\%-27\% less amount of effort than the baseline approach in order to find the same amount of actual defective lines.}
Figure~\ref{fig:rq4} shows that our approach achieves a median Effort@20\%Recall\textsubscript{line} of 0.20 for Openstack and 0.19 for Qt, while the baseline approach achieves a median Effort@20\%Recall\textsubscript{line} of 0.24 for OpenStack and 0.26 for Qt.
Similarly, our approach achieves a median IFA of 0 for OpenStack and 1 for Qt, while the baseline approach achieves a median IFA of 3 for OpenStack and 4 for Qt.
Our Wilcoxon signed-ranked test also confirms that the difference of Effort@20\%Recall\textsubscript{line} and IFA between our approach and the baseline is statistically significant ($p$-value $<$ 0.05) with a Cliff's $|\delta|$ effect size of large ($|\delta|=0.52-0.69$) for Effort@20\%Recall\textsubscript{line} and a Cliff's $|\delta|$ effect size of medium ($|\delta|=0.36-0.39$) for IFA.
% The statistical test confirms that our \name~approach can reduce the amount of effort by 17\%-20 when compared to the baseline approach in order to find the same amount of actual defective lines.

% achieves lower median IFA and Effort@20\%Recall, and achieves greater Top10-Accuracy and Recall\@20\%Effort when compared to N-gram approach at line-level ranking
% Fig \ref{fig:rq4} illustrates the evaluation result of Openstack and Qt, obtained from \name~and N-gram, of the following measures: IFA, Top10-Accuracy, Recall\@20\%Effort and Effort\@20\%Recall at line level.

% \todo{not sure how to explain IFA, please help me...}

% \todo{add statistical test result (if there is enough space}

% From the figure, our approach achieves 100\% greater Top10-Accuracy, 19-21\% higher Recall\@20\%Effort, and obtain 20-21\% lower Effort@20\%Recall than N-gram approach.
% Our approach also obtains the median IFA of zero, which is less than the value of N-gram.
% In particular, our approach can obtain more defective lines from a bug-introducing commit when observing 20\% of total LOC, while achieves greater accuracy than N-gram approach.
% Moreover, developers can spend around 20\% less effort to observe 20\% defective lines of the commit, when compared with the baseline.
% Thus, \name~can reduce developers' effort to inspect defective code lines better than N-gram approach.

%% file: sections/discussion.tex
\section{Discussion}

\subsection{Implications to Practitioners}

\emph{Our \name~approach may help practitioners to better prioritize defect-introducing commits and better identify defective lines,}
since we find that our \name~approach outperforms (RQ1), more cost-effective (RQ2), faster (RQ3), and more fine-grained (RQ4) than the state-of-the-art approaches (i.e., EALR, CC2Vec, and DeepJIT). 
Traditionally, Just-In-Time defect prediction methods only prioritize defect-introducing commits, saving a lot of code inspection effort. 
% However, we find that the ratio of actual defective lines for each commit is \todo{X\%-Y\%}.
However, we find that the average ratio of actual defective lines for each commit is 50\%.
Thus, developers still spend unnecessarily effort on inspecting clean lines.
In addition to predict defect-introducing commits, our \name~approach can also accurately predict defective lines within a defect-introducing commit, saving 17\%-20\% effort that developers need to spend when compared to the baseline approach~\cite{yan2020just}.

\subsection{Implications to Researchers}

\emph{Researchers should consider the key principles of Just-In-Time defect prediction models (i.e., to generate predictions as soon as possible),}
since the results of our replication study show that, when excluding testing datasets, the F-measure of CC2Vec approach is decreased by 38.5\% for OpenStack and 45.7\% for Qt.
In reality, it is unlikely that the unlabelled testing dataset would be available beforehand when training JIT models.
Thus, when conducting an experiment, testing data should be excluded when developing AI/ML models.

\emph{Researchers should explore simple solutions (i.e., Explainable AI approaches~\cite{jiarpakdee2021perception,jiarpakdee2020xai4se,tantithamthavorn2020explainable,wattanakriengkrai2020predicting,rajapaksha2021sqaplanner}) first over complex and compute-intensive deep learning approaches for SE tasks}, since we find that our \name~approach outperforms the deep learning approaches for Just-In-Time defect prediction.
This recommendation has been advocated by prior studies in other SE tasks~\cite{hellendoorn2017deep,fu2017easy,liu2018neural,menzies2018500+}.
For example, Menzies~\ea~\cite{menzies2018500+} suggested that researchers should explore simple and fast approaches before applying deep learning approaches on SE tasks.
Hellendoorn~\cite{hellendoorn2017deep} found that a careful implementation of NLP approaches outperform deep learning approaches.
Liu~\ea~\cite{liu2018neural} found that simple $k$-nearest neighbours approach outperforms neural machine translation approaches. 

%%%%%%%%%%%%%%%%%%%%%%%%%%%%%%%%
\subsection{Threats to Validity}\label{sec:threats}
%%%%%%%%%%%%%%%%%%%%%%%%%%%%%%%%

\emph{Threats to construct validity}  relates to the impact of parameter settings of the techniques that our approach relies upon (i.e, SMOTE, DE, Random Forest, and LIME)~\cite{tantithamthavorn2016automated, fu2016tuning, tantithamthavorn2018impact}.
To mitigate this threat, we apply a Differential Evolution algorithm to optimize the parameter setting of the SMOTE technique.
We use the parameter settings of DE, suggested by Agrawal~\ea~\cite{agrawal2018better}.
% We set the number of trees for Random Forest to 100, since we find that our approach yields similar performance when using other settings.
We use the default settings of LIME (i.e., the number of samples = 5,000).
For the baseline approaches, we use the best parameter settings provided by the implementation of the DeepJIT~\cite{hoang2019deepjit} and CC2Vec approaches~\cite{CC2Vec}.

Prior work raised concerns that the ground-truths data collection of defect-introducing commits could be delayed \cite{tan2015online,cabral2019class}. 
Thus, it is possible that our studied JIT datasets may be missing some of the false negative commits when defects are not fixed (i.e., defect-fixing commits that are not yet fixed). 
However, the goal of this paper is not to improve the data construction approach.
Instead, we use the same datasets that were used in the prior work for a fair comparison. 
Thus, future work should consider addressing this concern.

\emph{Threats to external validity} relates to the limited number of the studied datasets (i.e., OpenStack and Qt) to ensure a fair comparison with the CC2Vec approach~\cite{CC2Vec}.
Thus, other commit-level datasets can be explored in future work.

\emph{Threats to internal validity} relates to the randomization of several techniques that our approach relies upon~\cite{liem2020run}.
After we repeat our experiments with different random seeds, we observe minor differences (e.g., $\pm0.01$ for AUC).
Nevertheless, our JITLine approach is still the best performer for all RQs.
The used random seed number is reported in our replication package at Zenodo: \url{http://doi.org/10.5281/zenodo.4433498}.

% to foster the replicability.

We follow the experimental setting of the original study~\cite{hoang2019deepjit,CC2Vec} (i.e., one single training/testing data split without cross-validation). 
% Thus, we have only one performance value for each project.
% Therefore, statistical analysis and effect size analysis are not applied for RQ1, RQ2, and RQ3.
% by Oat
Therefore, statistical analysis and effect size analysis are not applied for RQ1, RQ2, and RQ3, since we have only one performance value for each project.
% However, we do our best

%% file: paper.bbl
% Generated by IEEEtranS.bst, version: 1.14 (2015/08/26)
\begin{thebibliography}{10}
\providecommand{\url}[1]{#1}
\csname url@samestyle\endcsname
\providecommand{\newblock}{\relax}
\providecommand{\bibinfo}[2]{#2}
\providecommand{\BIBentrySTDinterwordspacing}{\spaceskip=0pt\relax}
\providecommand{\BIBentryALTinterwordstretchfactor}{4}
\providecommand{\BIBentryALTinterwordspacing}{\spaceskip=\fontdimen2\font plus
\BIBentryALTinterwordstretchfactor\fontdimen3\font minus
  \fontdimen4\font\relax}
\providecommand{\BIBforeignlanguage}[2]{{%
\expandafter\ifx\csname l@#1\endcsname\relax
\typeout{** WARNING: IEEEtranS.bst: No hyphenation pattern has been}%
\typeout{** loaded for the language `#1'. Using the pattern for}%
\typeout{** the default language instead.}%
\else
\language=\csname l@#1\endcsname
\fi
#2}}
\providecommand{\BIBdecl}{\relax}
\BIBdecl

\bibitem{agrawal2019dodge}
A.~Agrawal, W.~Fu, D.~Chen, X.~Shen, and T.~Menzies, ``How to``dodge" complex
  software analytics,'' \emph{Transactions on Software Engineering (TSE)},
  2019.

\bibitem{agrawal2018better}
A.~Agrawal and T.~Menzies, ``Is ``better data" better than ``better data
  miners"?'' in \emph{Proceedings of the International Conference on Software
  Engineering (ICSE)}, 2018, pp. 1050--1061.

\bibitem{cabral2019class}
G.~G. Cabral, L.~L. Minku, E.~Shihab, and S.~Mujahid, ``Class imbalance
  evolution and verification latency in just-in-time software defect
  prediction,'' in \emph{Proceedings of the International Conference on
  Software Engineering (ICSE)}, 2019, pp. 666--676.

\bibitem{da2016framework}
D.~A. Da~Costa, S.~McIntosh, W.~Shang, U.~Kulesza, R.~Coelho, and A.~E. Hassan,
  ``A framework for evaluating the results of the szz approach for identifying
  bug-introducing changes,'' \emph{Transactions on Software Engineering (TSE)},
  pp. 641--657, 2016.

\bibitem{fu2017easy}
W.~Fu and T.~Menzies, ``Easy over hard: A case study on deep learning,'' in
  \emph{Proceedings of the Joint Meeting on Foundations of Software Engineering
  (FSE)}, 2017, pp. 49--60.

\bibitem{fu2016tuning}
W.~Fu, T.~Menzies, and X.~Shen, ``Tuning for software analytics: Is it really
  necessary?'' \emph{Information and Software Technology (IST)}, pp. 135--146,
  2016.

\bibitem{Fukushima2014}
T.~Fukushima, Y.~Kamei, S.~McIntosh, K.~Yamashita, and N.~Ubayashi, ``An
  empirical study of just-in-time defect prediction using cross-project
  models,'' in \emph{Proceedings of the Working Conference on Mining Software
  Repositories (MSR)}, 2014, pp. 172--181.

\bibitem{hellendoorn2017deep}
V.~J. Hellendoorn and P.~Devanbu, ``Are deep neural networks the best choice
  for modeling source code?'' in \emph{Proceedings of the Joint Meeting on
  Foundations of Software Engineering (FSE)}, 2017, pp. 763--773.

\bibitem{hoang2019deepjit}
T.~Hoang, H.~K. Dam, Y.~Kamei, D.~Lo, and N.~Ubayashi, ``{DeepJIT}: an
  end-to-end deep learning framework for just-in-time defect prediction,'' in
  \emph{Proceedings of the International Conference on Mining Software
  Repositories (MSR)}, 2019, pp. 34--45.

\bibitem{CC2Vec}
T.~Hoang, H.~J. Kang, D.~Lo, and J.~Lawall, ``{CC2Vec}: Distributed
  representations of code changes,'' in \emph{Proceedings of the International
  Conference on Software Engineering (ICSE)}, 2020, pp. 518--529.

\bibitem{huang2017supervised}
Q.~Huang, X.~Xia, and D.~Lo, ``Supervised vs unsupervised models: A holistic
  look at effort-aware just-in-time defect prediction,'' in \emph{Proceedings
  of the International Conference on Software Maintenance and Evolution
  (ICSME)}, 2017, pp. 159--170.

\bibitem{jiang2017automatically}
S.~Jiang, A.~Armaly, and C.~McMillan, ``Automatically generating commit
  messages from diffs using neural machine translation,'' in \emph{Proceedings
  of the International Conference on Automated Software Engineering (ASE)},
  2017, pp. 135--146.

\bibitem{jiarpakdee2021perception}
J.~Jiarpakdee, C.~Tantithamthavorn, and J.~Grundy, ``{Practitioners'
  Perceptions of the Goals and Visual Explanations of Defect Prediction
  Models},'' in \emph{Proceedings of the International Conference on Mining
  Software Repositories (MSR)}, 2021, p. To Appear.

\bibitem{jiarpakdee2020xai4se}
J.~Jiarpakdee, C.~Tantithamthavorn, H.~Khanh~Dam, and J.~Grundy, ``An empirical
  study of model-agnostics techniques for defect prediction models,''
  \emph{IEEE Transactions on Software Engineering (TSE)}, 2020.

\bibitem{jimenez2019importance}
M.~Jimenez, R.~Rwemalika, M.~Papadakis, F.~Sarro, Y.~Le~Traon, and M.~Harman,
  ``The importance of accounting for real-world labelling when predicting
  software vulnerabilities,'' in \emph{Proceedings of the Joint Meeting on
  European Software Engineering Conference and Symposium on the Foundations of
  Software Engineering (ESEC/FSE)}, 2019, pp. 695--705.

\bibitem{Kamei2010}
Y.~{Kamei}, S.~{Matsumoto}, A.~{Monden}, K.~{Matsumoto}, B.~{Adams}, and A.~E.
  {Hassan}, ``Revisiting common bug prediction findings using effort-aware
  models,'' in \emph{Proceedings of the International Conference on Software
  Maintenance (ICSM)}, 2010, pp. 1--10.

\bibitem{Kamei2013}
Y.~Kamei, E.~Shihab, B.~Adams, A.~E. Hassan, A.~Mockus, A.~Sinha, and
  N.~Ubayashi, ``A large-scale empirical study of just-in-time quality
  assurance,'' \emph{Transactions on Software Engineering (TSE)}, pp. 757--773,
  2012.

\bibitem{Kim2008}
S.~Kim, E.~J. Whitehead, and Y.~Zhang, ``Classifying software changes: clean or
  buggy?'' \emph{Transactions on Software Engineering (TSE)}, pp. 181--196,
  2008.

\bibitem{kim2007predicting}
S.~Kim, T.~Zimmermann, E.~J. Whitehead~Jr, and A.~Zeller, ``Predicting faults
  from cached history,'' in \emph{Proceedings of the International Conference
  on Software Engineering (ICSE)}, pp. 489--498.

\bibitem{Imblearn}
G.~Lema{{\^i}}tre, F.~Nogueira, and C.~K. Aridas, ``Imbalanced-learn: A python
  toolbox to tackle the curse of imbalanced datasets in machine learning,''
  \emph{Journal of Machine Learning Research}, pp. 1--5, 2017.

\bibitem{liem2020run}
C.~Liem and A.~Panichella, ``Run, forest, run? on randomization and
  reproducibility in predictive software engineering,'' \emph{arXiv preprint
  arXiv:2012.08387}, 2020.

\bibitem{liu2018neural}
Z.~Liu, X.~Xia, A.~E. Hassan, D.~Lo, Z.~Xing, and X.~Wang,
  ``Neural-machine-translation-based commit message generation: how far are
  we?'' in \emph{Proceedings of the International Conference on Automated
  Software Engineering (ASE)}, 2018, pp. 373--384.

\bibitem{McIntosh2018}
S.~McIntosh and Y.~Kamei, ``Are fix-inducing changes a moving target? a
  longitudinal case study of just-in-time defect prediction,''
  \emph{Transactions on Software Engineering (TSE)}, pp. 412--428, 2017.

\bibitem{mende2010effort}
T.~Mende and R.~Koschke, ``Effort-aware defect prediction models,'' in
  \emph{Proceedings of the European Conference on Software Maintenance and
  Reengineering (CSMR)}, 2010, pp. 107--116.

\bibitem{menzies2018500+}
T.~Menzies, S.~Majumder, N.~Balaji, K.~Brey, and W.~Fu, ``500+ times faster
  than deep learning:a case study exploring faster methods for text mining
  stackoverflow,'' in \emph{Proceedings of the International Conference on
  Mining Software Repositories (MSR)}, 2018, pp. 554--563.

\bibitem{PASCARELLA2019}
L.~Pascarella, F.~Palomba, and A.~Bacchelli, ``Fine-grained just-in-time defect
  prediction,'' \emph{Journal of Systems and Software}, pp. 22 -- 36, 2019.

\bibitem{rahman2019natural}
M.~Rahman, D.~Palani, and P.~C. Rigby, ``Natural software revisited,'' in
  \emph{Proceedings of the International Conference on Software Engineering
  (ICSE)}, 2019, pp. 37--48.

\bibitem{rajapaksha2021sqaplanner}
D.~Rajapaksha, C.~Tantithamthavorn, J.~Jiarpakdee, C.~Bergmeir, J.~Grundy, and
  W.~Buntine, ``{SQAPlanner: Generating Data-Informed Software Quality
  Improvement Plans},'' 2021.

\bibitem{Rajbahadur2017}
G.~K. Rajbahadur, S.~Wang, Y.~Kamei, and A.~E. Hassan, ``{The impact of using
  regression models to build defect classifiers},'' in \emph{Proceedings of the
  International Working Conference on Mining Software Repositories (MSR)},
  2017, pp. 135--145.

\bibitem{LIME}
M.~T. Ribeiro, S.~Singh, and C.~Guestrin, ``Why should {I} trust you?"
  explaining the predictions of any classifier,'' in \emph{Proceedings of the
  International Conference on Knowledge Discovery and Data Mining (SIGKDD)},
  2016, pp. 1135--1144.

\bibitem{rodriguez2018reproducibility}
G.~Rodr{\'\i}guez-P{\'e}rez, G.~Robles, and J.~M. Gonz{\'a}lez-Barahona,
  ``Reproducibility and credibility in empirical software engineering: A case
  study based on a systematic literature review of the use of the szz
  algorithm,'' \emph{Information and Software Technology (IST)}, pp. 164--176,
  2018.

\bibitem{Shivaji2013}
S.~Shivaji, E.~J. Whitehead, R.~Akella, and S.~Kim, ``Reducing features to
  improve code change-based bug prediction,'' \emph{Transactions on Software
  Engineering (TSE)}, pp. 552--569, 2012.

\bibitem{SZZalgo}
J.~{\'S}liwerski, T.~Zimmermann, and A.~Zeller, ``When do changes induce
  fixes?'' in \emph{Proceedings of the International Workshop on Mining
  Software Repositories (MSR)}, 2005, p. 1–5.

\bibitem{PyDriller}
D.~Spadini, M.~Aniche, and A.~Bacchelli, ``Pydriller: Python framework for
  mining software repositories,'' in \emph{Proceedings of the Joint Meeting on
  European Software Engineering Conference and Symposium on the Foundations of
  Software Engineering (ESEC/FSE)}, 2018, pp. 908--911.

\bibitem{Rainer1997}
R.~Storn and K.~Price, ``Differential evolution--a simple and efficient
  heuristic for global optimization over continuous spaces,'' \emph{Journal of
  Global Optimization}, pp. 341--359.

\bibitem{tan2015online}
M.~Tan, L.~Tan, S.~Dara, and C.~Mayeux, ``Online defect prediction for
  imbalanced data,'' in \emph{Proceedings of the International Conference on
  Software Engineering (ICSE)}, vol.~2, 2015, pp. 99--108.

\bibitem{tantithamthavorn2020impact}
C.~Tantithamthavorn, A.~E. Hassan, and K.~Matsumoto, ``The impact of class
  rebalancing techniques on the performance and interpretation of defect
  prediction models,'' \emph{Transactions on Software Engineering (TSE)}, 2020.

\bibitem{tantithamthavorn2020explainable}
C.~Tantithamthavorn, J.~Jiarpakdee, and J.~Grundy, ``{Explainable AI for
  Software Engineering},'' \emph{arXiv preprint arXiv:2012.01614}, 2020.

\bibitem{tantithamthavorn2016automated}
C.~Tantithamthavorn, S.~McIntosh, A.~E. Hassan, and K.~Matsumoto, ``Automated
  parameter optimization of classification techniques for defect prediction
  models,'' in \emph{Proceedings of the International Conference on Software
  Engineering (ICSE)}, 2016, pp. 321--332.

\bibitem{tantithamthavorn2018impact}
------, ``The impact of automated parameter optimization on defect prediction
  models,'' \emph{Transactions on Software Engineering (TSE)}, pp. 683--711,
  2018.

\bibitem{SciPy}
P.~Virtanen, R.~Gommers, T.~E. Oliphant, M.~Haberland, T.~Reddy, D.~Cournapeau,
  E.~Burovski, P.~Peterson, W.~Weckesser, J.~Bright \emph{et~al.}, ``Scipy 1.0:
  fundamental algorithms for scientific computing in python,'' \emph{Nature
  methods}, pp. 261--272, 2020.

\bibitem{wattanakriengkrai2020predicting}
S.~Wattanakriengkrai, P.~Thongtanunam, C.~Tantithamthavorn, H.~Hata, and
  K.~Matsumoto, ``Predicting defective lines using a model-agnostic
  technique,'' \emph{Transactions on Software Engineering (TSE)}, 2020.

\bibitem{yan2020just}
M.~Yan, X.~Xia, Y.~Fan, A.~E. Hassan, D.~Lo, and S.~Li, ``Just-in-time defect
  identification and localization: A two-phase framework,'' \emph{Transactions
  on Software Engineering (TSE)}, 2020.

\bibitem{yang2016effort}
Y.~Yang, Y.~Zhou, J.~Liu, Y.~Zhao, H.~Lu, L.~Xu, B.~Xu, and H.~Leung,
  ``Effort-aware just-in-time defect prediction: simple unsupervised models
  could be better than supervised models,'' in \emph{Proceedings of the
  International Symposium on Foundations of Software Engineering (FSE)}, 2016,
  pp. 157--168.

\end{thebibliography}
